\documentclass[aps,prd,preprintnumbers,superscriptaddress,tighten,nofootinbib]{revtex4}
\usepackage{graphicx}
\usepackage{epsfig}
\usepackage{amsmath,color}

\begin{document}

\title{Reducing $\theta_{13}$ to $9^\circ$}

\author{Werner Rodejohann}
\email{werner.rodejohann@mpi-hd.mpg.de}

\affiliation{Max-Planck-Institut f{\"u}r Kernphysik, Postfach
103980, 69029 Heidelberg, Germany}

\author{He Zhang}
\email{he.zhang@mpi-hd.mpg.de}

\affiliation{Max-Planck-Institut f{\"u}r Kernphysik, Postfach
103980, 69029 Heidelberg, Germany}

\date{\today}

\begin{abstract}
\noindent We propose to consider the possibility that the observed value of $\theta_{13}$ is
not the result of a correction from an initially vanishing value, but rather the result of a
correction from an initially {\it larger} value. As an explicit example of this approach, we
consider analytically and numerically well-known CKM-like charged lepton corrections to a neutrino
diagonalization matrix that corresponds to a certain mixing scheme. Usually this results in
generating $\theta_{13} = 9^\circ$ from zero. We note here, however, that 9 is not only given by $0
+ 9$, but also by $18 - 9$. Hence, the extreme case of an initial value of 18 degrees,
{\it reduced} by charged lepton corrections to 9 degrees, is possible.
For some cases under study
new sum rules for the mixing parameters, and correlations with CP phases are found.
\end{abstract}

\maketitle

\section{Introduction}
\label{sec:intro}

Remarkable experimental activity in the past decades has
established that the phenomenon of neutrino flavor transition is described by neutrino
oscillations. Recent measurements of the smallest mixing angle $\theta_{13}$ at
reactor~\cite{An:2012eh,Ahn:2012nd,Abe:2012tg,An:2012bu}
and accelerator~\cite{Abe:2013xua} neutrino experiments have finally led
to an emerging picture where the order of magnitude of all elements of the PMNS matrix is
known. Theorists now face the task to understand and/or explain that structure.
Most flavor symmetry models \cite{Altarelli:2010gt,Ishimori:2010au,King:2013eh} were
constructed when only an upper limit on $\theta_{13}$ was known, and therefore
aimed at explaining $\theta_{13}=0$. Corrections to generate a non-zero value are then
applied.
In the present paper we depart from the historically motivated approaches to generate
non-zero $\theta_{13}$ from an initially vanishing value, and
consider the possibility that initially $\theta_{13}$ is already large. Now the usual
corrections to model predictions can reduce the initial value of $\theta_{13}$ to its observed
value. Of course, the phenomenology will then be different from the standard case.
As an explicit example on the consequences that follow, we consider charged lepton corrections.

No matter if neutrinos are Majorana or Dirac particles, the lepton flavor
mixing matrix stems from the mismatch between the diagonalization of the
charged lepton mass matrix $m_\ell$ and the neutrino mass matrix $m_\nu$, i.e.
\begin{eqnarray}\label{eq:main}
U=U^\dagger_\ell U_\nu \; ,
\end{eqnarray}
where $U_\ell$ and $U_\nu$ are the unitary matrices diagonalizing $m_\ell$ and $m_\nu$,
respectively. Now one can apply the following strategy to generate non-zero $\theta_{13} = \arcsin
|U_{e3}|$. Assuming that $(U_\nu)_{13}=0$, as well as $(U_\nu)_{23}=(U_\nu)_{33}$, and that
$U_\ell$ is related to the CKM matrix, i.e.\ essentially the unit matrix except for $(U_\ell)_{12}
= \lambda = \sin \theta_{\rm C}$, it follows that $|U_{e3}| = \lambda/\sqrt{2}$, or $\theta_{13}  =
9^\circ = 0 + 9^\circ$. Numerically, this is basically the observed value of about $\theta_{13} =
9^\circ$, and the fact that this lepton mixing parameter is numerically
connected to quark parameters seems
to support this argument, but is of course not a proof\footnote{The observed value of $|U_{e3}|$ is
also close to $\sqrt{m_s/m_b}$, which is just a coincidence.}.  Nevertheless, relating the
charged lepton diagonalization to the CKM matrix can be arranged in grand unified models,
especially based on $SU(5)$, for which $m_\ell=m^T_d$ is a typical outcome. Such a relation has to
be viewed as an approximation due to the distinct mass spectra of leptons and quarks, and is
modified by higher order corrections or Clebsch-Gordon coefficients.  Nevertheless, models
predicting $U_{\rm CKM} \simeq U_\ell$ have been constructed, which in addition have
$(U_\nu)_{13}=0
$~\cite{Marzocca:2011dh,Antusch:2012fb,Meroni:2012ty,Antusch:2013kna,Marzocca:2013cr}. Hence, the
above strategy to generate $|U_{e3}| = \lambda/\sqrt{2}$, where $\lambda \simeq \sin\theta_{\rm C}
\simeq 0.23$, is based on actual model building foundations.
We will use for the sake of simplicity and definiteness $U_{\rm CKM} = U_\ell$ in what follows.

While the relation $9^\circ = 0 + 9^\circ$ has its virtues and attraction, one should not ignore
the possibility that $9^\circ = 18^\circ - 9^\circ$. This means that initially $U_{\nu}$ contains a
too large value of its 13-element, which is {\it reduced} to its observed value by a sizable
charged lepton correction, a CKM-like one in our case. Since the remaining lepton mixing angles are
necessarily non-zero both in $U$ and in $U_\ell$, the question arises whether $\theta_{13}$ should
initially be non-zero in the first place. This so far overlooked possibility is what we investigate
here, by performing a general analysis of Eq.\ (\ref{eq:main}) when $U_\ell$ is fixed to the CKM
matrix. The case of initially vanishing $(U_\nu)_{13}=0$ has been analyzed countless times, but the
cases when $|(U_\nu)_{13}| \simeq |U_{e3}|$, or more interestingly $|(U_\nu)_{13}| > |U_{e3}|$,
have never been considered. As a result we find new interesting sum rules, and also note the
already mentioned extreme case of reducing $\theta_{13}$ from 18 degrees to 9 degrees, where the
initial value could be obtained from flavor symmetries, as $18^\circ = \pi/10$ is related to
symmetries of geometrical objects.

The remainder of this paper is organized as follows. In Sec.~\ref{sec:II}, we present the general
formalism and derive the charged lepton corrections to an arbitrary $U_\nu$. Interesting sum rules
between neutrino mixing parameters are summarized. In Sec.~\ref{sec:numerics}, a detailed numerical
analysis of the model parameters and predictions is performed. Finally, in Sec.~\ref{sec:summary},
we state our conclusions.

\section{Methodology}
\label{sec:II}

In the picture of three-flavor neutrino oscillations, the lepton
flavor mixing is described by a $3\times 3$ unitary matrix $U$,
which is conventionally parametrized by three mixing angles
($\theta_{12}$, $\theta_{23}$ and $\theta_{13}$), and three CP
violating phases out of which one is the Dirac phase ($\delta$) and
the other two are the Majorana phases ($\rho$ and $\sigma$). In the
standard parametrization, the lepton mixing matrix is given by
\begin{widetext}
\begin{eqnarray}\label{eq:SP}
U = \left(\begin{matrix}c_{12} c_{13} & s_{12} c_{13} & s_{13}
e^{-{\rm i}\delta} \cr -s_{12} c_{23} - c_{12} s_{23} s_{13}e^{{\rm
i}\delta} & c_{12} c_{23} - s_{12} s_{23} s_{13}e^{{\rm i}\delta} &
s_{23} c_{13} \cr s_{12} s_{23} - c_{12} c_{23} s_{13}e^{{\rm
i}\delta} & -c_{12} s_{23} - s_{12} c_{23} s_{13}e^{{\rm i}\delta} &
c_{23} c_{13}
\end{matrix}\right) \left(\begin{matrix} e^{{\rm i}\rho} & 0 & 0
\cr 0 & e^{{\rm i}\sigma} & 0 \cr 0 & 0 & 1 \end{matrix}\right)  ,
\end{eqnarray}
\end{widetext}
where $s_{ij} \equiv \sin \theta_{ij}$ and $c_{ij} \equiv \cos
\theta_{ij}$ (for $ij = 12, 23, 13$). In case of Dirac neutrinos the
phases $\rho$ and $\sigma$ will be irrelevant. The results of this paper are independent on
the nature of the neutrino. The latest
global analysis of current neutrino oscillation data yields
\cite{GonzalezGarcia:2012sz}
\begin{eqnarray}\label{eq:bound}
\sin^2\theta_{12} = 0.313^{+0.013}_{-0.012} \; , \nonumber \\
\sin^2\theta_{23} = 0.444^{+0.036}_{-0.031} \; , \\
\sin^2\theta_{13} = 0.0244^{+0.0020}_{-0.0019} \; , \nonumber
\end{eqnarray}
where short baseline reactor data with baseline shorter than 100~m are not included. Another
recent fit result is obtained in \cite{Capozzi:2013csa}, with similar results. There are also
non-trivial results on the CP phase $\delta$, with best-fit results around $3\pi/2$, or $\cos
\delta \simeq  0$. However, the $1\sigma$ ranges are very large, including essentially also the
case $\cos \delta \simeq -1$. We note that for some of the cases that we will discuss it is
actually crucial whether $\cos \delta$ is 0 or $-1$, and therefore we use only the obtained ranges
of the mixing angles in our fits.

The concrete form of $U_\ell$ cannot be fixed unless a specific mode is considered. Motivated by the connection between the CKM matrix and $U_\ell$ in many
grand unified models we assume here for definiteness $U_\ell=U_{\rm CKM}$.
As for the unitary matrix $U_\nu$ diagonalizing the neutrino
mass matrix, one can parametrize it in analogy to $U$ by using three rotation angles $\tilde\theta_{12}$, $\tilde\theta_{23}$, and
$\tilde\theta_{13}$ together with a phase $\phi$. Note that we have ignored the Majorana-like  phases in this parametrization, since they are located on
the right-hand-side of $U_\nu$ and hence do not affect our discussions on the mixing angles and Dirac CP phase. Now the lepton flavor mixing matrix
is given by\footnote{For the case that $U_\nu$ is CKM-like, see \cite{Hochmuth:2007wq}.}
\begin{eqnarray}\label{eq:U}
U=U^\dagger_{\rm CKM} P
U_\nu(\tilde\theta_{12},\tilde\theta_{23},\tilde\theta_{13},\phi) \; .
\end{eqnarray}
Here $P={\rm diag}(e^{{\rm i}x},e^{{\rm i}y},1)$ is a phase matrix stemming from the
mismatch between $U_{e}$ and $U_{\nu}$ \cite{Frampton:2004ud}.

We proceed to expand the mixing matrix $U$ in order to obtain the charged lepton corrections.
Different from the lepton sector, the CKM matrix takes a nearly diagonal form, and is typically
parametrized by using four parameters ($\lambda$, $A$, $\rho$ and $\eta$) in the Wolfenstein
parametrization. Since we are mainly interested in the lepton flavor mixing which has not been
measured as precisely as $U_{\rm CKM}$, we will keep the Wolfenstein parametrization only up to
$\lambda^2$, i.e.
\begin{eqnarray}\label{eq:CKM}
U_{\rm CKM} \simeq \begin{pmatrix} 1-\frac{1}{2}\lambda^2 & \lambda
& 0 \cr -\lambda & 1-\frac{1}{2}\lambda^2 & A\lambda^2 \cr 0 &
-A\lambda^2 & 1
\end{pmatrix}  .
\end{eqnarray}
Now, by inserting Eq.~\eqref{eq:CKM} into \eqref{eq:U} we obtain the matrix elements of
$U$ to order $\lambda$ as\footnote{Ignoring CP phases,
expressions for the PMNS mixing angles in case of CKM-like corrections to
$U_\nu$, with angles in $U_\nu$ all larger than the ones in $U_\ell$ can be found in \cite{Ohlsson:2005js}.}
\begin{eqnarray}
U_{e1} & = & \tilde c_{12} \tilde c_{13} +\left( \tilde s_{12} \tilde c_{23} e^{-{\rm i}\varphi} + \tilde s_{23} \tilde c_{12} \tilde s_{13} e^{-{\rm
i}(\varphi-\phi)}\right) \lambda \; ,\label{eq:Ue1} \\
U_{e2} & = & \tilde s_{12} \tilde c_{13} +\left( -\tilde c_{12} \tilde c_{23} e^{-{\rm i}\varphi} + \tilde s_{23} \tilde s_{12} \tilde s_{13} e^{-{\rm i}(\varphi-\phi)}\right) \lambda  \; , \label{eq:Ue2}\\
\label{eq:Ue3} U_{e3} & = &  \tilde s_{13} e^{-{\rm i} \phi} - \tilde s_{23} \tilde c_{13} e^{-{\rm i}\varphi} \lambda \; , \\
U_{\mu 3} & = &  \tilde s_{23} \tilde  c_{13} + \tilde s_{13} e^{{\rm i}(\varphi-\phi)} \lambda  \; ,\label{eq:Um3}
\end{eqnarray}
where $\varphi=x-y$ has been defined, and the notation $\tilde s_{ij} \equiv
\sin\tilde\theta_{ij}$, $\tilde c_{ij} \equiv \cos\tilde\theta_{ij}$ is adopted. Since the
charged lepton mixing matrix takes the CKM form, only the $12$-rotation plays a role. Consequently,
one can rotate away one of the phases, leaving only the difference between two CP phases $x$ and
$y$ in the above results.

Comparing with the standard parametrization given in Eq.~\eqref{eq:SP}, we find
\begin{eqnarray}\label{eq:t13}
\sin^2\theta_{13} & \simeq &  \tilde s^2_{13} -2 \lambda \tilde s_{13} \tilde c_{13} \tilde s_{23} \cos(\varphi-\phi) + \lambda^2 (\tilde s^2_{23}
\tilde c^2_{13} - \tilde s^2_{13}) \; ,  \\ \label{eq:t12}
\sin^2\theta_{12} & \simeq & \tilde s^2_{12} -2 \lambda \frac{1}{\tilde c_{13}} \tilde s_{12} \tilde c_{12} \tilde c_{23} \cos \varphi \; ,
 \\
\sin^2\theta_{23} & \simeq & \tilde s^2_{23} + 2 \lambda \frac{1}{\tilde c_{13}} \tilde s_{23} \tilde s_{13} \tilde c^2_{23} \cos(\varphi-\phi) \; .
\label{eq:t23}
\end{eqnarray}
where the ${\cal O}(\lambda^2)$ terms are only kept for $\sin^2\theta_{13}$, since $\theta_{13}$ is
relatively smaller compared to the other mixing angles. 
As for the Dirac phase $\delta$, to leading order we have
\begin{eqnarray}\label{eq:delta}
\tan\delta = \frac{ \tilde s_{23} \tilde c_{13} \lambda s_\varphi -\tilde s_{13} s_\phi}{\tilde s_{13}c_\phi - \tilde s_{23} \tilde c_{13} \lambda c_\varphi} \; ,
\end{eqnarray}
where $s_\phi =\sin\phi$, $s_\varphi = \sin \varphi$ and so on. It might also be useful to express the Jarlskog
invariant~\cite{Jarlskog:1985ht,Wu:1985ea} in terms of the model parameters, i.e.
\begin{eqnarray}\label{eq:Jcp}
J_{\rm CP} & = &  \tilde J_{\rm CP} + \lambda \tilde c_{13} \tilde
c_{23} \left\{ \tilde s_{12} \tilde c_{12} (\tilde c^2_{23}-\tilde
c^2_{13})\sin \varphi \right. \nonumber \\
&& \left. +\tilde s_{13} \tilde s_{23} \left[ \tilde c_{23} (\tilde
c^2_{12} -\tilde s^2_{12}) \sin(\varphi-\phi)-\tilde s_{12} \tilde
s_{23} \tilde s_{13} \tilde c_{12} \sin(\varphi-2\phi) \right] \right\}
 \nonumber \\
& \simeq & \tilde J_{\rm CP} + \lambda \tilde s_{12} \tilde c_{12} \tilde c_{13} \tilde
c_{23}    (\tilde c^2_{23}-\tilde
c^2_{13})\sin \varphi
\end{eqnarray}
where, as usual, $\tilde J_{\rm CP}$ is defined as
\begin{eqnarray}\label{eq:Jcpt}
\tilde J_{\rm CP} = \tilde s_{12} \tilde s_{23} \tilde s_{13} \tilde
c_{12} \tilde c_{23} \tilde c^2_{13} \sin\phi \; .
\end{eqnarray}
Of course, even if $\tilde \theta_{13}=0$ is assumed, CP violation can still be induced by
the $\lambda$ correction, when $\sin \varphi=\sin(x-y)\neq 0$.

Both $\theta_{13}$ and $\theta_{23} $ are independent of $\tilde \theta_{12}$ at leading order. The
leading corrections to $\theta_{13}$ and $\theta_{23} $ are proportional to $\lambda \, \tilde
s_{13}$, whereas the leading correction to $\theta_{12}$ is proportional to $\lambda$. This
indicates that a larger deviation of $\tilde \theta_{12}$ from $\theta_{12}$ than for the other
mixing angles is allowed. However, there are terms including cosines of phases in the expressions,
which can suppress the corrections. Note that the same combination of phases appears in the
expressions for $\sin^2\theta_{23}$ and $\sin^2\theta_{13}$, which implies a correlation between
both observables, if the second order term in  $\sin^2\theta_{13}$ can be ignored. It reads
\begin{eqnarray}\label{eq:sum0}
\sin^2\theta_{23} - \sin^2\tilde\theta_{23}
&=&-\frac{\cos^2\tilde\theta_{23}}{\cos^2\tilde\theta_{13} }
\left( \sin^2\theta_{13} - \sin^2\tilde\theta_{13}\right).
\end{eqnarray}
In case $\phi=0$, there is a correlation between the 23- and 12-sectors:
\begin{eqnarray}\label{eq:sum1}
\sin^2\theta_{23} - \sin^2\tilde\theta_{23}
&=&-\frac{\cos\tilde\theta_{23}\sin\tilde\theta_{23}\sin\tilde\theta_{13}}
{\cos\tilde\theta_{12} \sin\tilde\theta_{12} }
\left( \sin^2\theta_{12} - \sin^2\tilde\theta_{12}\right).
\end{eqnarray}
However, the general case is complicated and depends on many parameters.
The obvious extreme cases are $\tilde \theta_{13}=0$, $\tilde \theta_{13}> \theta_{13}$ and
$\tilde \theta_{13} \simeq \theta_{13}$. We will in the following discuss
these cases analytically, before performing a general numerical analysis.

\subsection{The case of $\tilde \theta_{13} =0 $}
\label{sec:IIA}

We will start from the most simple case with $\tilde \theta_{13}=0$, though there is nothing new
too add to existing knowledge (see e.g.\
\cite{Giunti:2002ye,Giunti:2002pp,Romanino:2004ww,Altarelli:2004jb,Minakata:2004xt,Frampton:2004ud,Petcov:2004rk,Masina:2005hf,King:2005bj,Plentinger:2005kx,Ohlsson:2005js,Antusch:2005kw,Hochmuth:2006xn,Goswami:2009yy,Dev:2011bd,Acosta:2014dqa}).
In the limit under study, the expressions for the mixing angles reduce to leading order to
\begin{eqnarray}
\sin\theta_{13} & \simeq & \lambda \sin\theta_{23} \; , \nonumber \\
\delta & \simeq & \varphi+\pi \; , \nonumber \\
\sin^2\theta_{12} & \simeq & \tilde s^2_{12} -2 \lambda \tilde s_{12} \tilde c_{12} \tilde c_{23} \cos \varphi \; .
\end{eqnarray}
From the relation $\sin\theta_{13} \simeq\lambda \sin\tilde \theta_{23} $ one obtains for
$\tilde\theta_{23}=\pi/4$ the value $\sin^2\theta_{13} \simeq 0.0255$, in very good agreement with
the measured value. In the tri-bimaximal mixing case, we have
\begin{eqnarray}
\sin\theta_{23} & = & \frac{1}{\sqrt{2}} \; , \nonumber \\
\sin\theta_{13} & = & \frac{1}{\sqrt{2}}\lambda  \; , \nonumber \\
\delta & = & \varphi+\pi \; ,\nonumber  \\
\sin^2\theta_{12} & = & \frac{1}{3} +\frac{2\sqrt{2}}{3} \sin\theta_{13} \cos \delta \; ,\nonumber
\end{eqnarray}
whereas for the bimaximal mixing case we obtain
\begin{eqnarray}
\sin\theta_{23} & = & \frac{1}{\sqrt{2}} \; , \nonumber \\
\sin\theta_{13} & = & \frac{1}{\sqrt{2}}\lambda  \; , \nonumber \\
\delta & = & \varphi+\pi \; , \nonumber \\
\sin^2\theta_{12} & = & \frac{1}{2} + \sin\theta_{13} \cos \delta \; .\nonumber
\end{eqnarray}
In the tri-bimaximal based case, $\delta$ has to be close to $\pi/2$ (or $3\pi/2$) in order to
suppress the $\theta_{13}$ correction to $\sin^2\theta_{12}=1/3$. The situation is however
different in the bimaximal case, in which a sizable and negative $\theta_{13}$-correction is
required in order to reduce the maximal mixing value $\sin^2\tilde\theta_{12}=1/2$. Hence, $\delta
\simeq \pi$ or $2\pi$ has to be fulfilled.
This interplay of the mixing scheme (bimaximal/tri-bimaximal) in $U_\nu$ and the Dirac phase
in neutrino oscillations has first been noticed in \cite{Plentinger:2005kx}. Recall that the fit
results from Ref.\ \cite{Capozzi:2013csa,GonzalezGarcia:2012sz} include at $1\sigma$ essentially
both cases, $\delta \simeq 2\pi$ and $\delta \simeq 3\pi/2$, where the latter value is close
to the best-fit one.

\subsection{The case of $\tilde \theta_{13}> \theta_{13}$}
\label{sec:IIB}

If $\tilde \theta_{13}$ is larger than the observed value of $\theta_{13}$,
the term proportional to $\lambda^2$ term in Eq.~\eqref{eq:t13} can be neglected,
leaving us with a set of novel sum rules. Appealing values of the initial value are
e.g.\ $\tilde \theta_{13}=\pi/10$ or $\tilde \theta_{13}=\pi/12$. Assuming $\tilde \theta_{13}=\pi/10$
 (or $\tilde \theta_{13} =18^\circ$) and
for simplicity also $\tilde \theta_{23}=\pi/4$, the following sum rules can be deduced:
\begin{eqnarray}
\sin^2\theta_{13} & \simeq & \frac{3-\sqrt{5}}{8} -\frac{(\sqrt{5}-1)\sqrt{5+\sqrt{5}}}{8} \lambda \cos(\varphi-\phi) \; , \\
\sin^2\theta_{23} & \simeq & \frac{1}{2} -\frac{4}{5+\sqrt{5}} \left(\sin^2\theta_{13}-\frac{3-\sqrt{5}}{8}\right)  ,  \label{eq:30}\\
\sin^2\theta_{12} & \simeq & \sin^2\tilde\theta_{12} -\frac{2}{\sqrt{5+\sqrt{5}}} \lambda \sin2\tilde \theta_{12}\cos \varphi\; .
\end{eqnarray}
Thus, using the measured value $\theta_{13} \simeq 9^\circ$ and Eq.~\eqref{eq:30}, one predicts
$\theta_{23} \simeq 47.3^\circ$.
Another interesting example is $\tilde \theta_{13}=\pi/12$ (or $\tilde \theta_{13} =15^\circ$),
which leads to the following rum rules,
\begin{eqnarray}
\sin^2\theta_{13} & \simeq & \frac{2-\sqrt{3}}{4} -\frac{\sqrt{2}}{4} \lambda \cos(\varphi-\phi) \; , \label{eq:21}\\
\sin^2\theta_{23} & \simeq & \frac{1}{2} - 2(2-\sqrt{3})\left(\sin^2\theta_{13}-\frac{2-\sqrt{3}}{4}\right)  , \label{eq:23} \\
\sin^2\theta_{12} & \simeq & \sin^2\tilde\theta_{12} - (\sqrt{3}-1) \lambda \sin2\tilde \theta_{12}\cos \varphi \; .
\end{eqnarray}
By inserting $\theta_{13} = 9^\circ$ into Eq.~\eqref{eq:23} we obtain the prediction
$\theta_{23} \simeq 46.3^\circ$. As in the previous example, we find $\theta_{23}$ in the second octant.

It is obvious from Eq.\ (\ref{eq:delta}) or from (\ref{eq:Ue1}$-$\ref{eq:Um3}) that in case
$(\tilde U_\nu)_{e3} > U_{e3}$ at leading order $\delta \simeq \phi$ holds. In addition, from
 (\ref{eq:21}) it is clear that the first and second terms should cancel to a large
extent in order to reduce to the observed value of $|U_{e3}|^2$.
To this end, the cosine in (\ref{eq:21}) should be close to 1,
which gives
\begin{eqnarray} \label{eq:25}
\delta \simeq \phi \simeq \varphi\; .
\end{eqnarray}
Similar to the discussion in the previous subsection, if $\sin^2\tilde \theta_{12} =1/3 $ holds,
$\delta \simeq \pi/2$ (or $3\pi/2$)
is required to suppress its corrections to $\theta_{12}$. In contrast, for
$\sin^2\tilde \theta_{12} =1/2$, $\delta \simeq \pi$ is expected in order to avoid a too large
solar mixing angle. Amusingly, the correlation between $\sin^2\tilde
\theta_{12}$ and CP violation is identical to the one for vanishing $\tilde \theta_{13}$.
Both cases can in principle
be distinguished by their prediction for $\theta_{23}$, see the blue and red
points in the lower left plot in Fig.\ \ref{fig:5}.

\subsection{The case of $\tilde \theta_{13} \simeq \theta_{13}$, or
$\sin \tilde \theta_{13}  \simeq \sin\tilde\theta_{23} \lambda$} \label{sec:IIC}

This is obviously the most complicated case, and does not allow much analytical results.
The Dirac CP phase is determined by
\begin{eqnarray}
\delta = - {\rm Arg} ( \tilde s_{13} e^{-{\rm i} \phi} - \tilde s_{23} \tilde c_{13} e^{-{\rm i}\varphi} \lambda) \; .
\end{eqnarray}
or by Eq.~\eqref{eq:delta}.
In principle, any value for $\delta$ is possible. As an interesting example, we look at
the scenario with $\tilde \theta_{13} =9^\circ$ (or $\tilde \theta_{13} =\pi/20$).
In this special case, the sum of the first and third term of
Eq.~\eqref{eq:t13} is about 0.05, the same size as the second term if the cosine would not be there.
Since the measured $\theta_{13}$ is also very close to $9^\circ$, one would naturally expect that the
phase difference between $\phi$ and $\varphi$ is around $\pm \pi/3$.
Concretely, we have the following relation
\begin{eqnarray} \label{eq:27}
\delta \simeq  \phi \pm \pi/3 \simeq \varphi \pm 2\pi/3 \; .
\end{eqnarray}
Note also that corrections to $\theta_{12}$ are not sensitive to $\tilde \theta_{13}$ as shown in
the general formula \eqref{eq:t12}, which implies that the CP phase $\delta$ is restricted to be
close to $\pm \pi/6$ and $\pm \pi/3$ for $\tilde s^2_{12}=1/3$ and $\tilde s^2_{12}=1/2$,
respectively.

\section{Numerics}
\label{sec:numerics}

In this section we fit the five parameters ($\tilde \theta_{12}$, $\tilde \theta_{23}$, $\tilde
\theta_{13}$, $\phi$ and $\varphi$) to the experimental data using the exact form of
Eq.~\eqref{eq:U}. To figure out the allowed parameter spaces of the model parameters, we compare
the latest global-fit data with a $\chi^2$-function defined as
\begin{eqnarray}\label{eq:chi2}
\chi^2_{ij} = \sum_{i<j} \frac{(\sin^2\theta_{ij} -
\sin^2\theta^0_{ij})^2}{\sigma^2_{ij}} \ ,
\end{eqnarray}
where $\theta^0_{ij}$ represents the experimental data given in
Eq.~\eqref{eq:bound}, $\sigma_{ij}$ denote the corresponding 1$\sigma$
absolute errors, and $\theta_{ij}$ are the predictions of the model and
can be expressed in terms of the model parameters.

\subsection{$\tilde \theta_{12}$-$\tilde \theta_{13}$ plane}

\begin{figure}[h]
\includegraphics[width=0.4\textwidth]{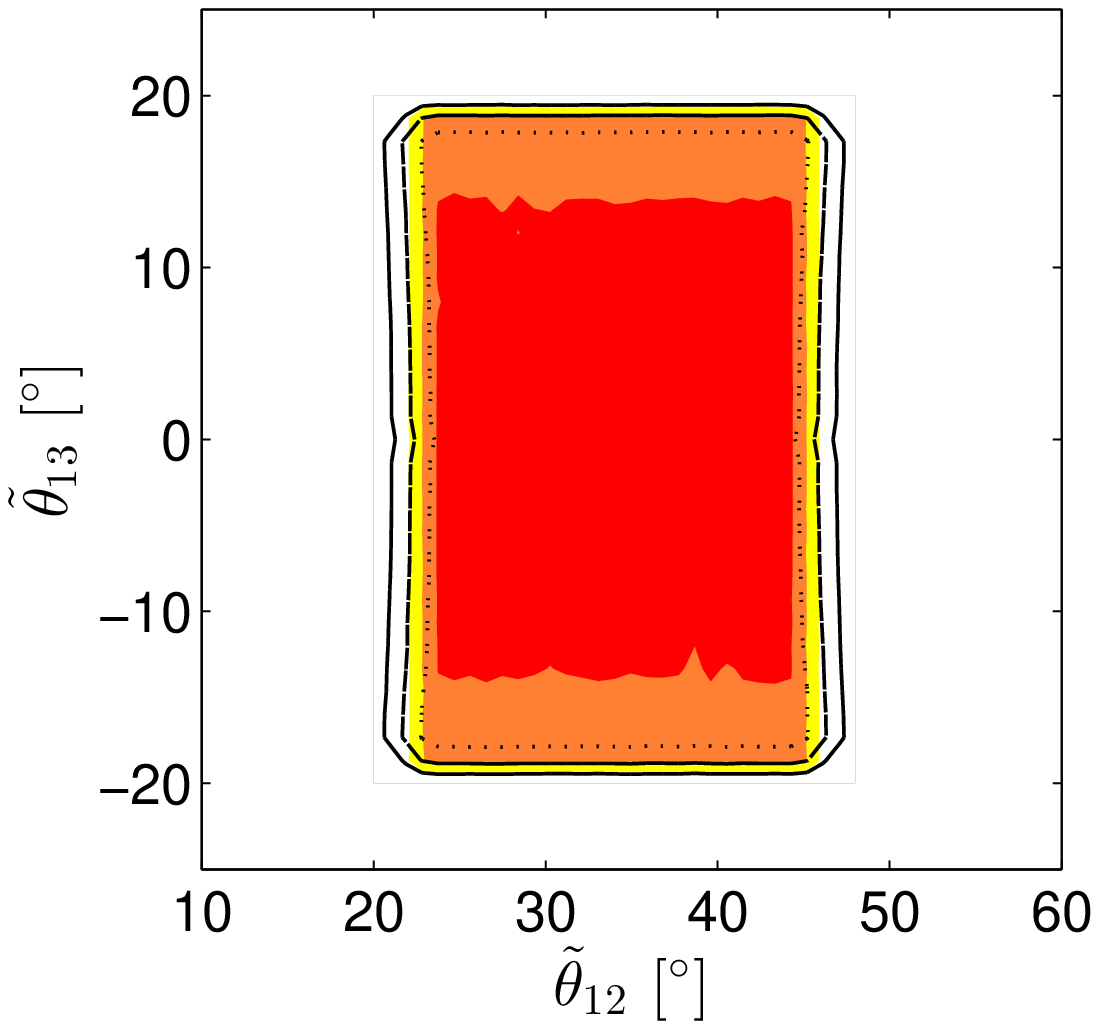}\vspace{-0.0cm}  \\
\includegraphics[width=0.4\textwidth]{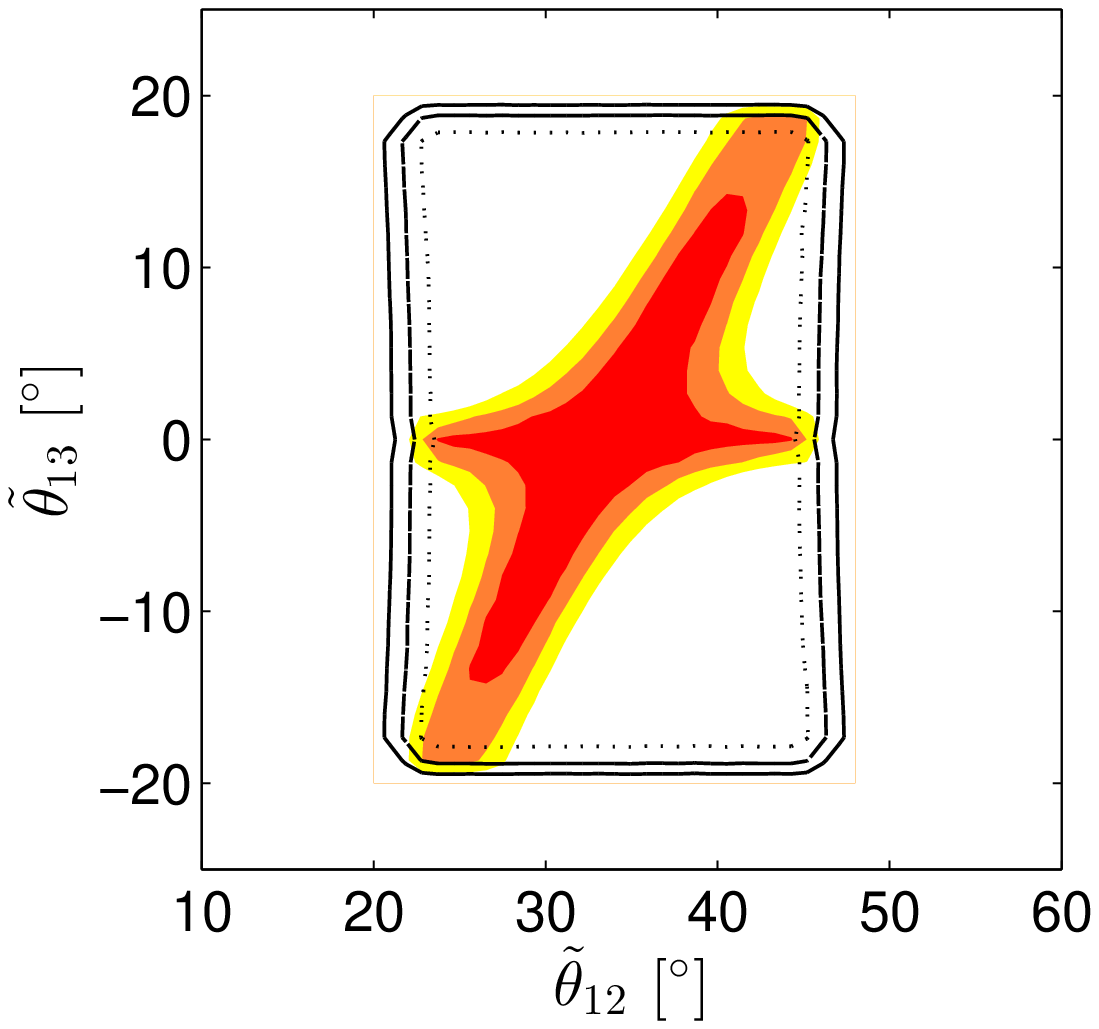}\vspace{-0.0cm} \\
\includegraphics[width=0.4\textwidth]{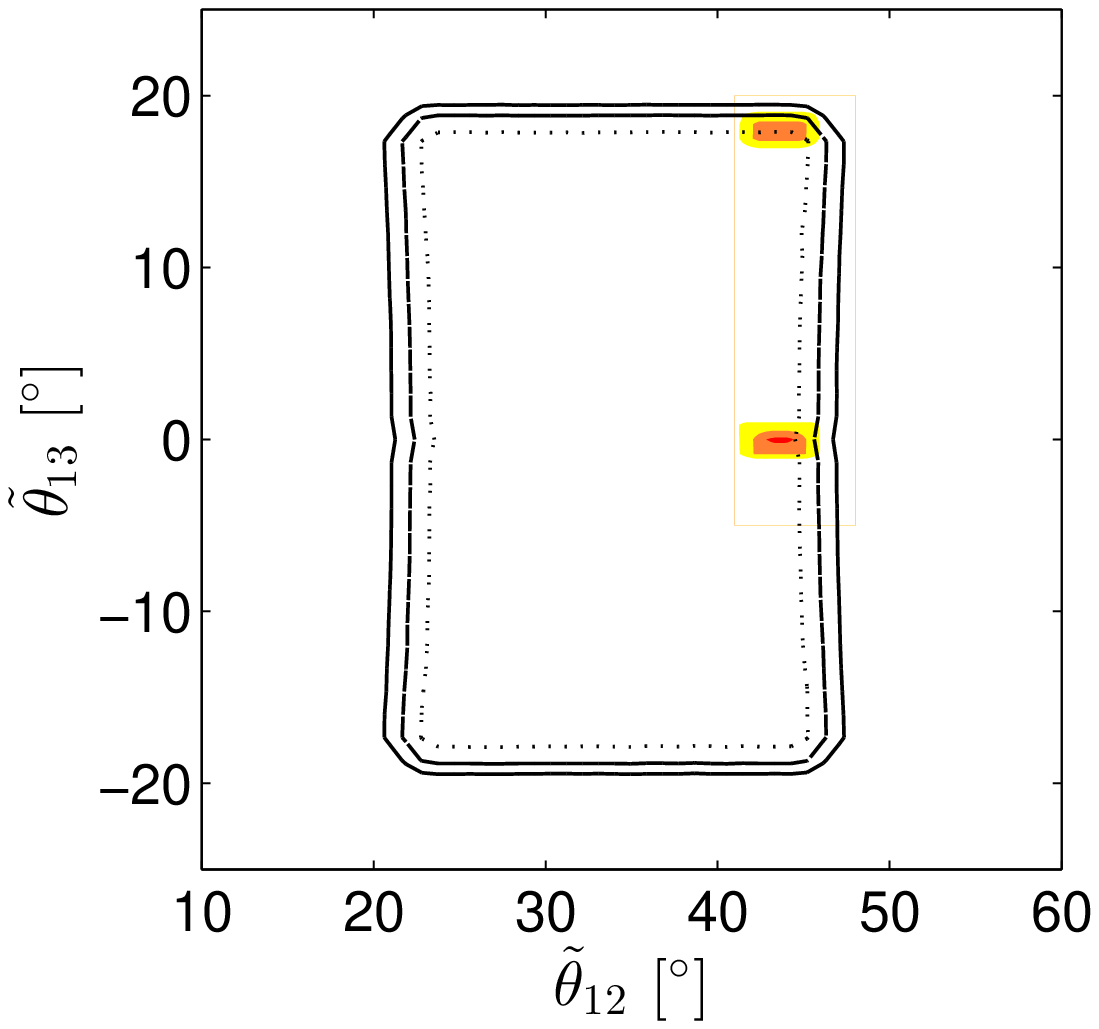}\vspace{-0cm}
\caption{\label{fig:1} Parameter ranges of $\tilde \theta_{12}$ and $\tilde \theta_{13}$ at 1, 2
and 3$\sigma$. For the color contours, we have fixed $\tilde \theta_{23}=45^\circ$. In the upper panel, we allow all phases to freely vary between $0$ and $2\pi$. In the middle panel, we switch off $\phi$ but not $\varphi$, whereas in the lower panel, all CP phases are set to zero.}
\end{figure}

We start from projecting the parameter space to the $\tilde \theta_{12}$-$\tilde \theta_{13}$ plane.
The parameter ranges for $\tilde \theta_{12}$ and $\tilde \theta_{13}$
are shown in Fig.~\ref{fig:1} using contour lines for the most general case.
We also consider the case of maximal $\tilde \theta_{23}$ using colored contours, and make assumptions
about the CP phases.

From Fig.~\ref{fig:1} we see that $\tilde \theta_{13}$ can be as large as $19.2^\circ$, which
inspires us with mixing patterns such as $\sin^2(\pi/10)=(3-\sqrt{5})/8$ and $\sin^2({\pi}/{12})=(2-\sqrt{3})/4$. Such values of $\pi$ divided by $n$ can be obtained
in flavor symmetry models such as in Refs.\ \cite{Blum:2007nt,Adulpravitchai:2009bg}. The range of
$\tilde \theta_{12}$ is wide and a maximal $\tilde \theta_{12}$ can be accommodated. If $\tilde
\theta_{23}$ is fixed to $\pi/4$, the parameter space shrinks only slightly, which is a consequence
of the suppressed (by both $\lambda$ and $\tilde \theta_{13}$) correction terms to $\tilde
\theta_{23}$, see Eq.~\eqref{eq:t23}. In the limit $\phi=0$, for which the 12- and 13 sectors are
correlated, see Eq.~\eqref{eq:sum1}, a sizable $\tilde \theta_{13}$ demands a relatively large
value of $\cos \varphi$ in order to suppress its contribution to $\theta_{13}$. This in turn
requires $\tilde \theta_{12}$ to be close to maximal. In contrast, if $\tilde \theta_{13}$ is tiny,
the constraint on $\tilde \theta_{12}$ becomes less stringent, which can be seen clearly from our analytical
results Eq.~\eqref{eq:t12}. Explicitly, for a vanishing $\tilde \theta_{13}$, one has the
approximate relation $\sin\theta_{13} \simeq \lambda \sin\theta_{23} $. In such a case, the leading
order correction to $\tilde \theta_{12}$ is flexible since it is proportional to $\cos \varphi$. If
all phases are zero, a significant and negative correction to $\tilde \theta_{12}$ is expected, and
consequently only the nearly maximal value $\tilde \theta_{12} \simeq \pi/4$ can be accommodated.

\subsection{$\tilde \theta_{12}$--$\tilde \theta_{23}$ plane}

\begin{figure}
\includegraphics[width=0.4\textwidth]{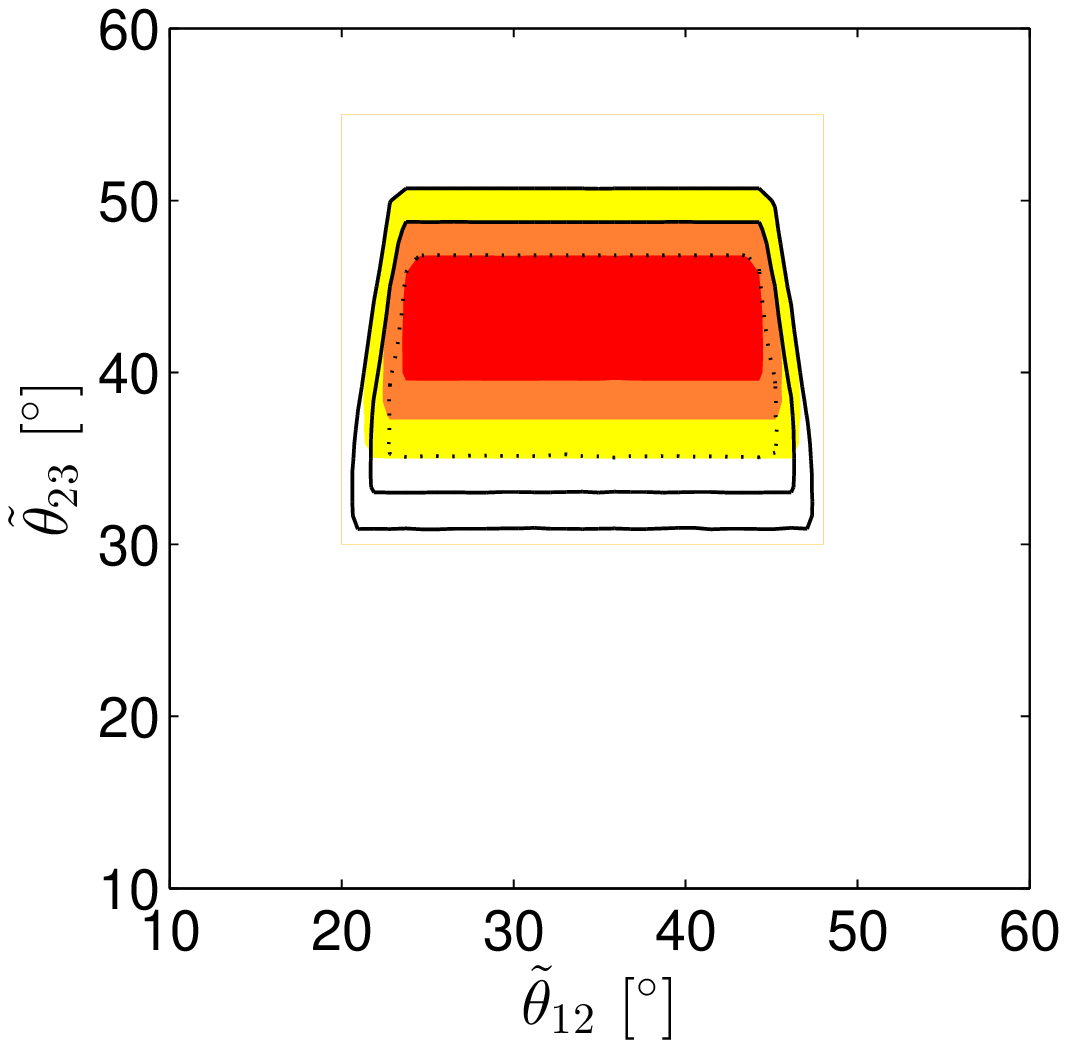}\vspace{-0.0cm}
\includegraphics[width=0.4\textwidth]{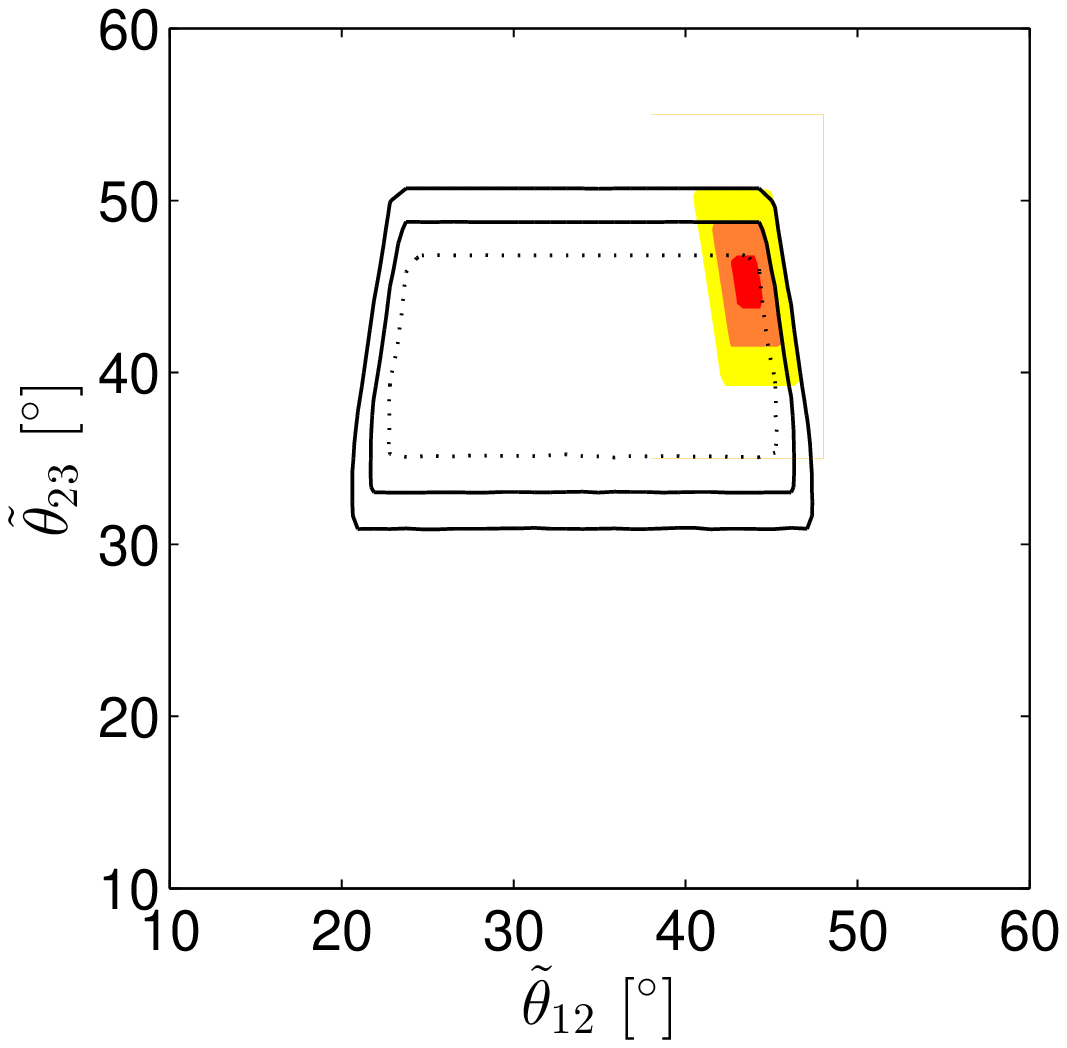}\vspace{-0.0cm} \\
\includegraphics[width=0.4\textwidth]{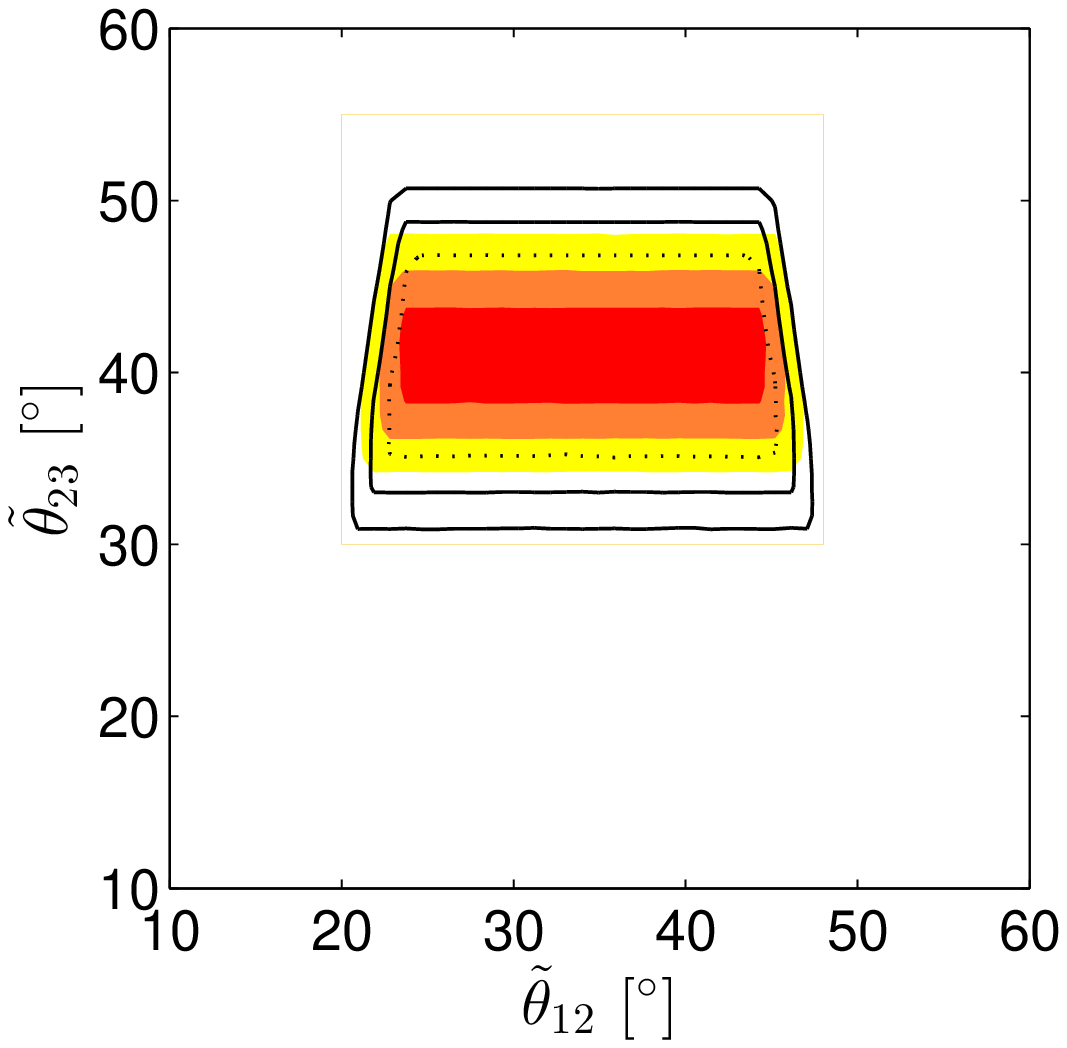}\vspace{-0.0cm}
\includegraphics[width=0.4\textwidth]{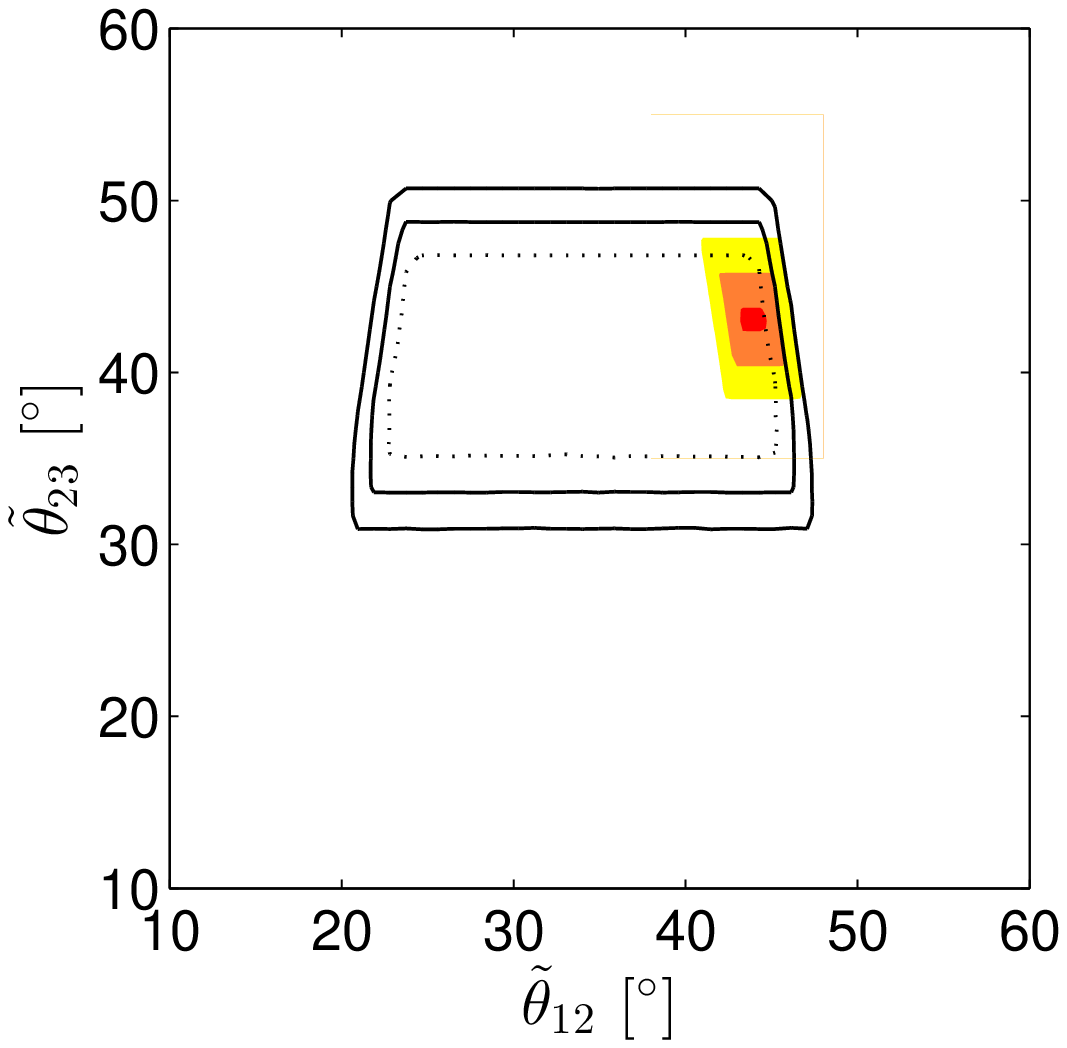}\vspace{-0.0cm}
\caption{\label{fig:2} The parameter ranges of $\tilde \theta_{12}$ and $\tilde \theta_{23}$ at 1, 2
and 3$\sigma$. For the color contours $\tilde \theta_{13}=0$ (upper row) or
$\tilde \theta_{13} = \pi/10$ (lower row) is fixed, but the choices of the phases are different.
In the left column, we allow all phases to freely vary between $0$ and $2\pi$,
whereas in the right column, all phases are set to zero.}
\end{figure}

The allowed parameter space in the $\tilde \theta_{12}$--$\tilde \theta_{23}$ plane is
shown in Fig.~\ref{fig:2}.
As special cases, we choose $\tilde \theta_{13}=0$ and $\tilde \theta_{13} = \pi/10$, both for the
general case and for all phases being set to zero.

As expected from the suppressed corrections to $\tilde \theta_{23}$,
the parameter range of $\tilde \theta_{23}$ is similar to that of $\theta_{23}$.
If we neglect the CP phases, $\tilde \theta_{13}=0$ leads to a large negative correction to
$\tilde \theta_{12}$, and a relatively larger $\tilde \theta_{12}$ is favored.
In case of large $\tilde \theta_{13}$, $\tilde \theta_{23}$ is driven towards smaller values, see
Eq.\ (\ref{eq:sum0}).

\subsection{$\tilde \theta_{13}$--$\tilde \theta_{23}$ plane}

\begin{figure}
\includegraphics[width=0.4\textwidth]{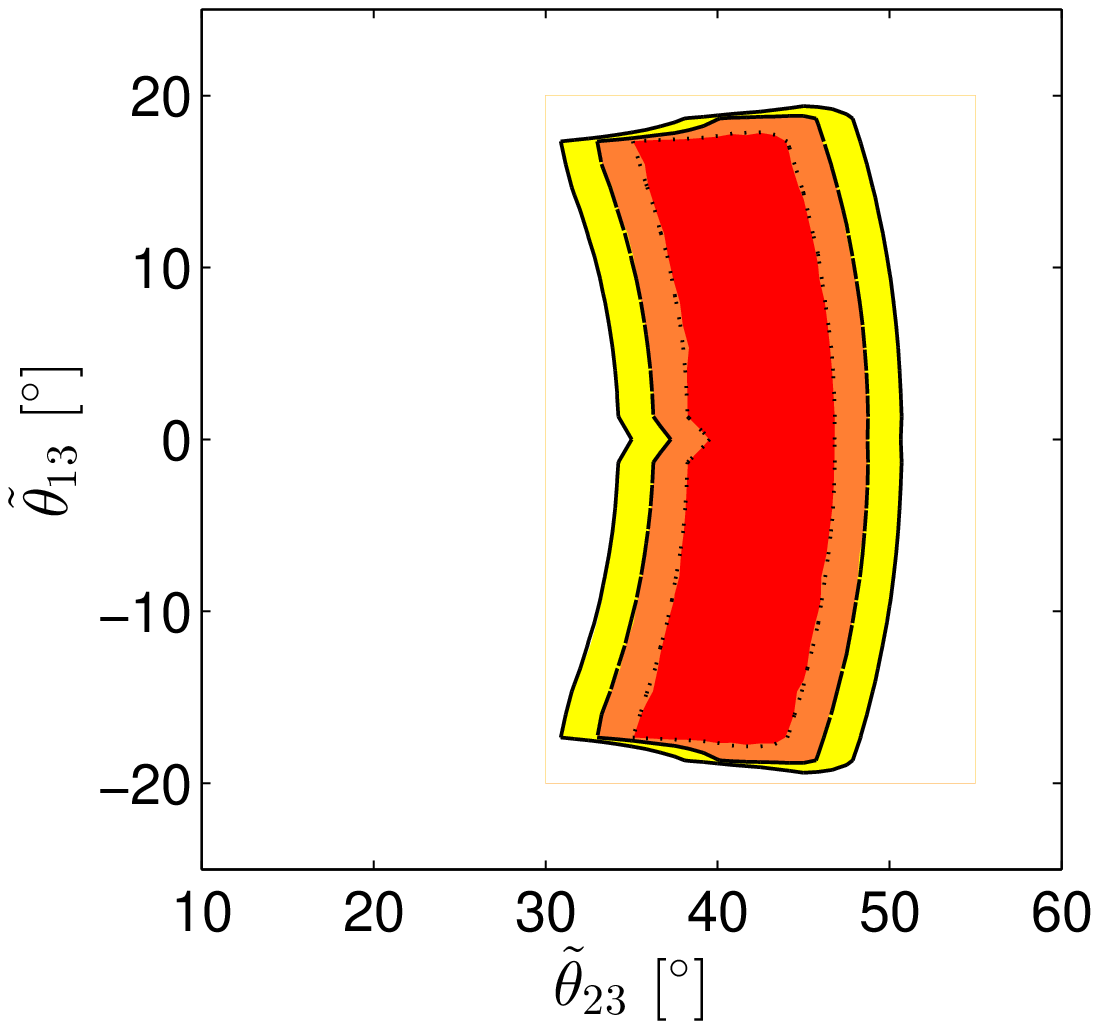}\vspace{-0.0cm}
\includegraphics[width=0.4\textwidth]{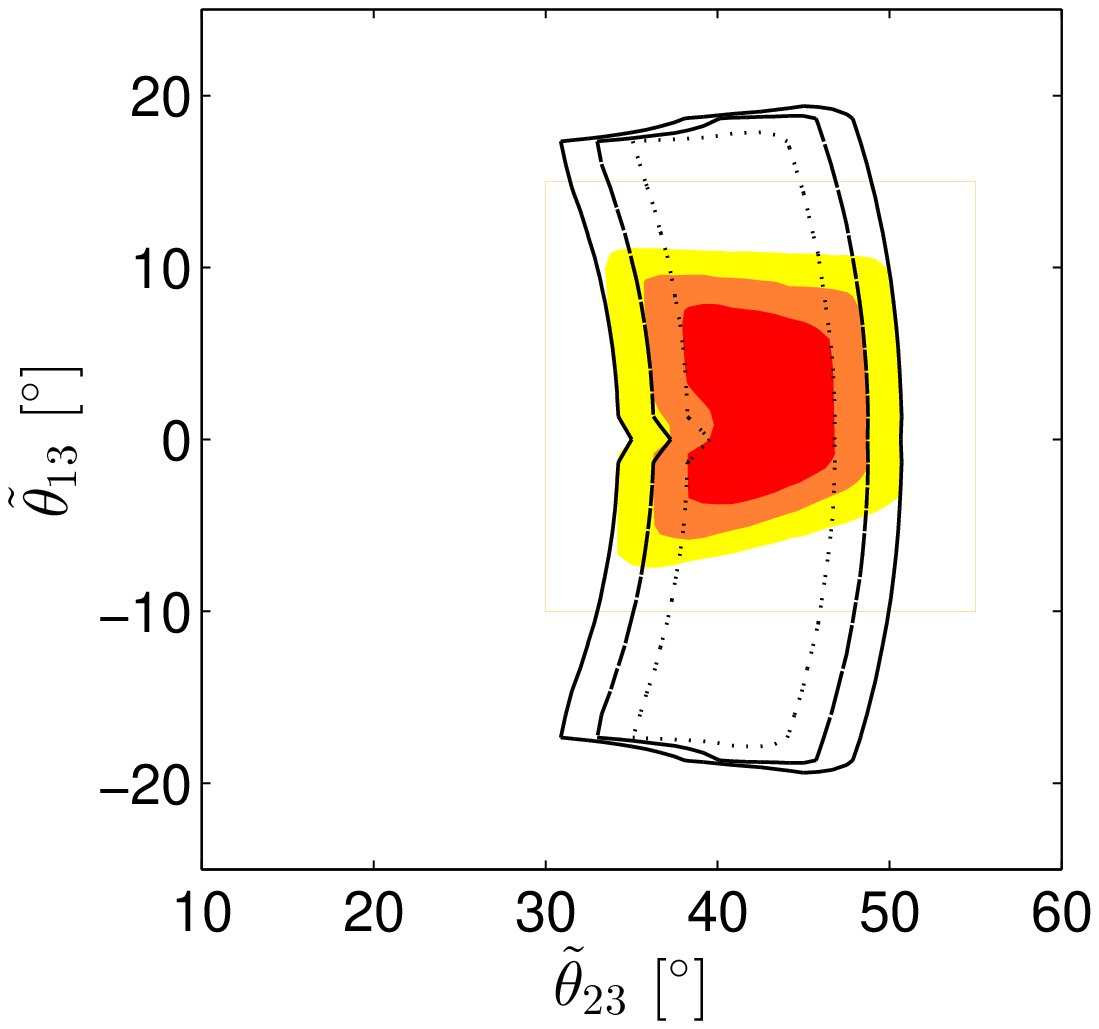}\vspace{-0.0cm} \\
\includegraphics[width=0.4\textwidth]{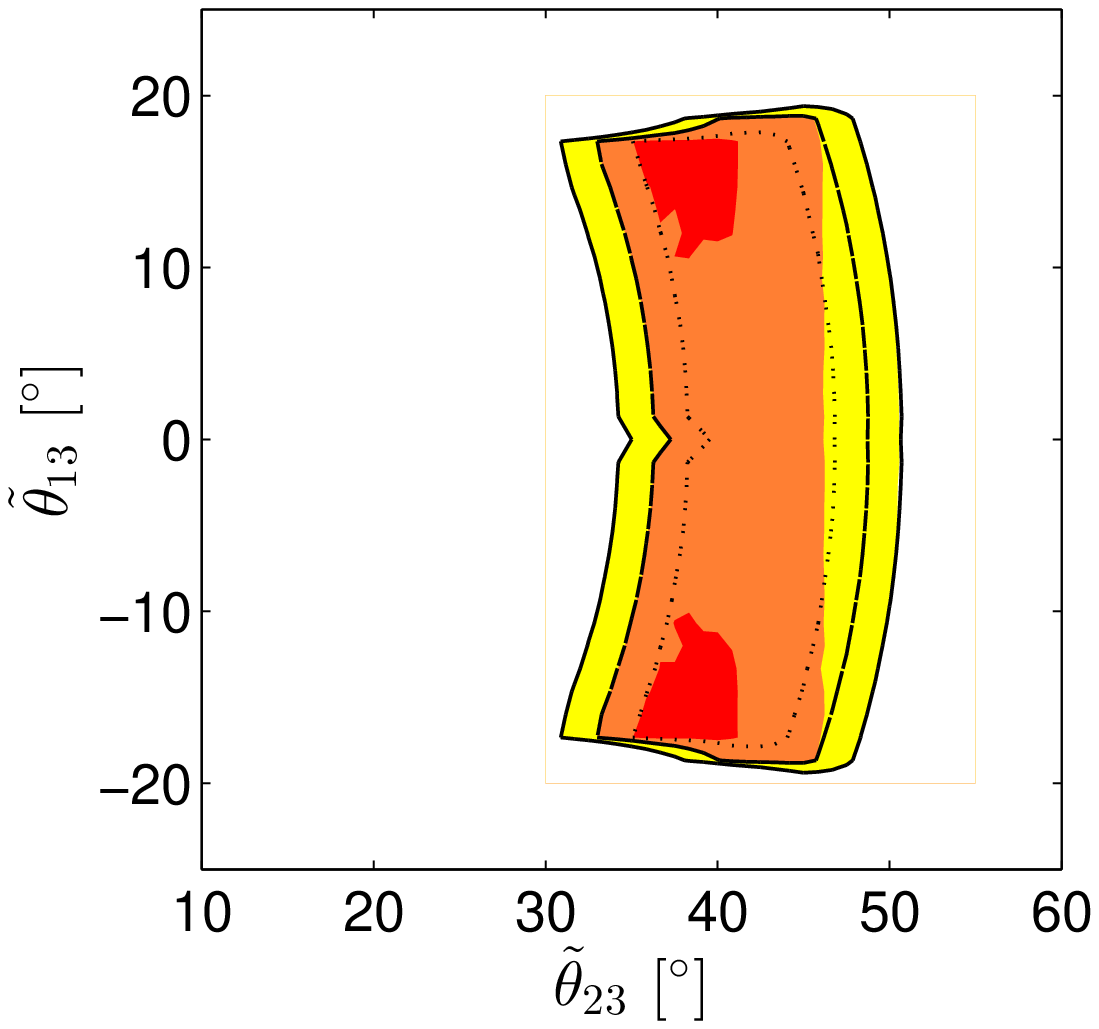}\vspace{-0.0cm}
\includegraphics[width=0.4\textwidth]{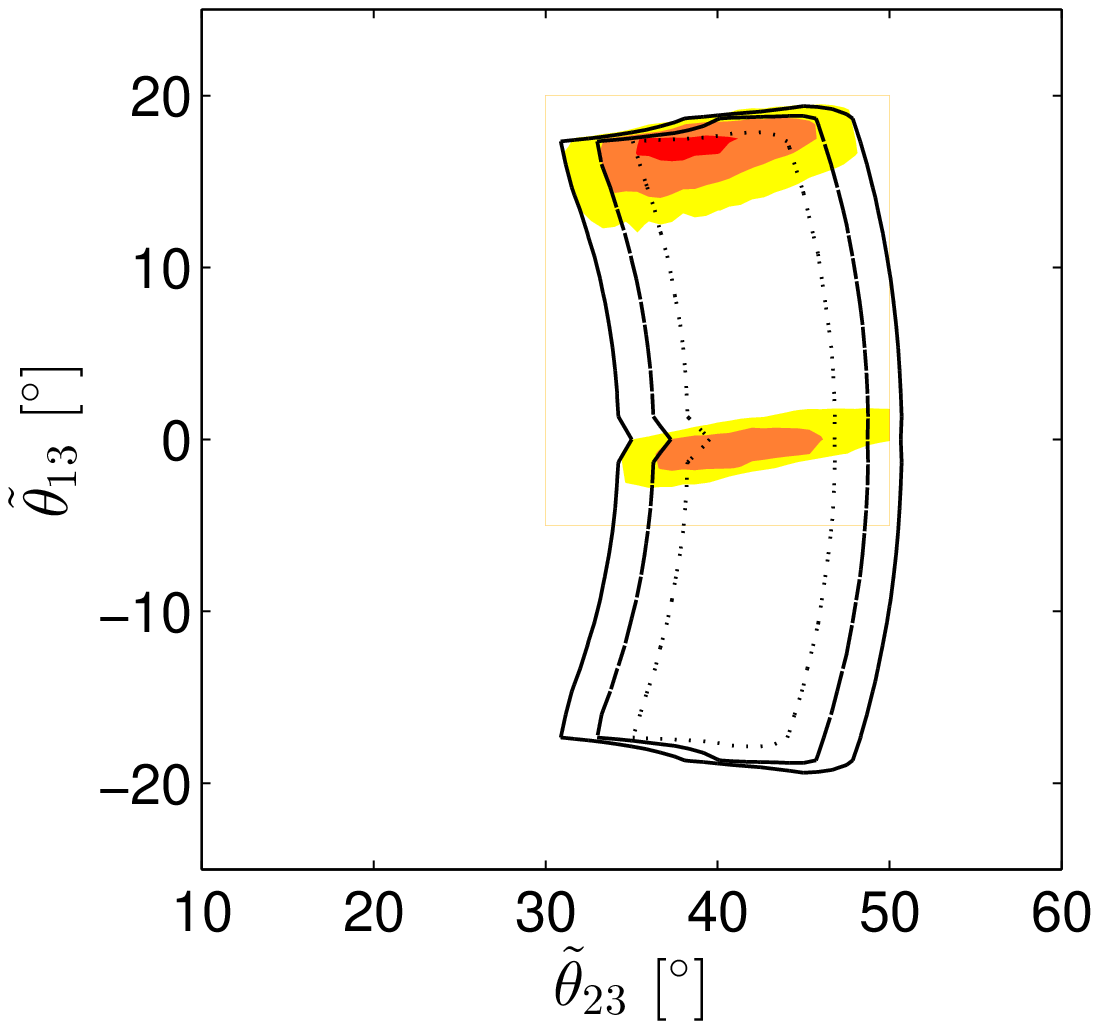}\vspace{-0.0cm}
\caption{\label{fig:3} The parameter range of $\tilde \theta_{13}$ and $\tilde \theta_{23}$ at 1, 2
and 3$\sigma$. For the color contours, we fix $\sin^2\tilde \theta_{12}=1/3$ in the upper row and
$\sin^2\tilde \theta_{12}=1/2$ in the lower row. In the left column, we allow all
 phases to freely vary between $0$ and $2\pi$, whereas in the right column
$\phi=0$ is fixed.}
\end{figure}

The allowed parameter space in the $\tilde \theta_{13}$--$\tilde \theta_{23}$ plane is shown
in Fig.~\ref{fig:3}.
As special cases we choose $\sin^2\tilde\theta_{12}=1/3$ and $\sin^2\tilde \theta_{12}=1/2$.

As the figure shows, $\tilde \theta_{13}$ and $\tilde \theta_{23}$ are not sensitive to the choice
of $\tilde \theta_{12}$, which has already been shown in the analytical part above, cf.\ Eqs.\
(\ref{eq:t13}, \ref{eq:t23}). They are however very sensitive to the CP phases, i.e.\ $\phi=0$
restricts the range of $\tilde \theta_{13}$ down to $-10^\circ \lesssim \tilde \theta_{13} \lesssim
10^\circ$ in the case of $\sin^2\tilde \theta_{12}=1/3$, and in two distinct regions around $0$ and
$18^\circ$ in the case of $\sin^2\tilde \theta_{12}=1/2$, with $\tilde\theta_{13} \sim 9^\circ$
being excluded. It is worth noting that, when all the phases are set to zero, there is no
parameter space for $\sin^2\tilde \theta_{12}=1/3$, since the derived $\theta_{12}$ is too small.

\subsection{$\varphi$--$\tilde \theta_{12}$ plane}

\begin{figure}[b]
\includegraphics[width=0.4\textwidth]{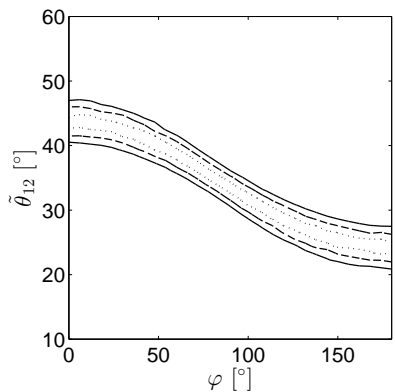}
\caption{\label{fig:4} The parameter range of $\varphi$ and $\tilde \theta_{12}$ at 1, 2 and
3$\sigma$. All other model parameters are marginalized.}
\end{figure}

As pointed out in the analytical section, the phase difference $\varphi=x-y$ is very crucial for
certain mixing patterns, in particular for $\tilde \theta_{12}$.
Thus, we illustrate the relation between $\varphi$ and $\tilde \theta_{12}$
in Fig.~\ref{fig:4}.
The correlation between small phases for $\sin^2 \tilde \theta_{12} = 1/2$ and
phases around $\pi$ for $\sin^2 \tilde \theta_{12} = 1/3$ is reproduced. Note that this feature is
present for all values of $\tilde \theta_{13}$.

\subsection{$J_{\rm CP}$--$\tilde \theta_{12}$ plane}

\begin{figure}
\includegraphics[width=0.4\textwidth]{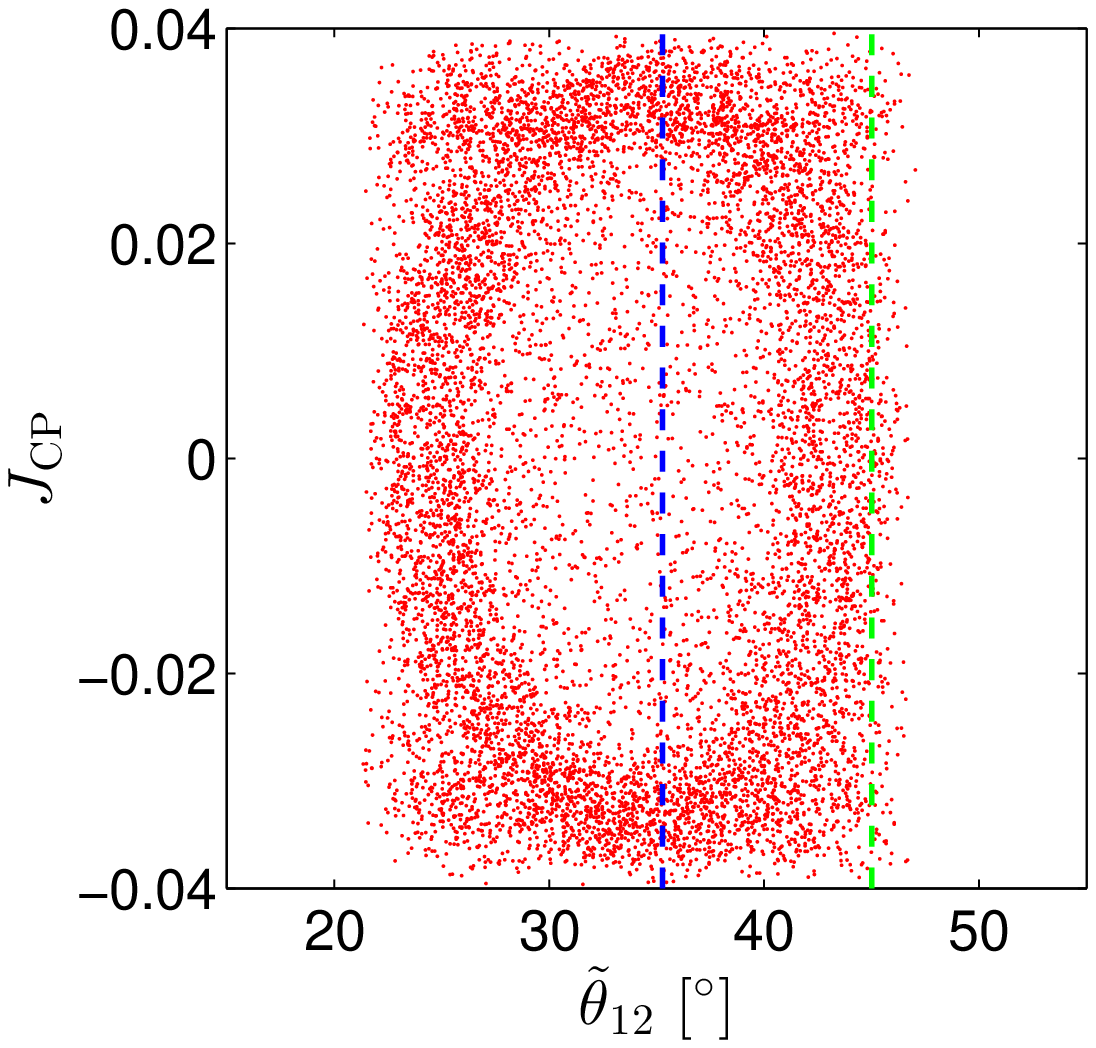}
\includegraphics[width=0.4\textwidth]{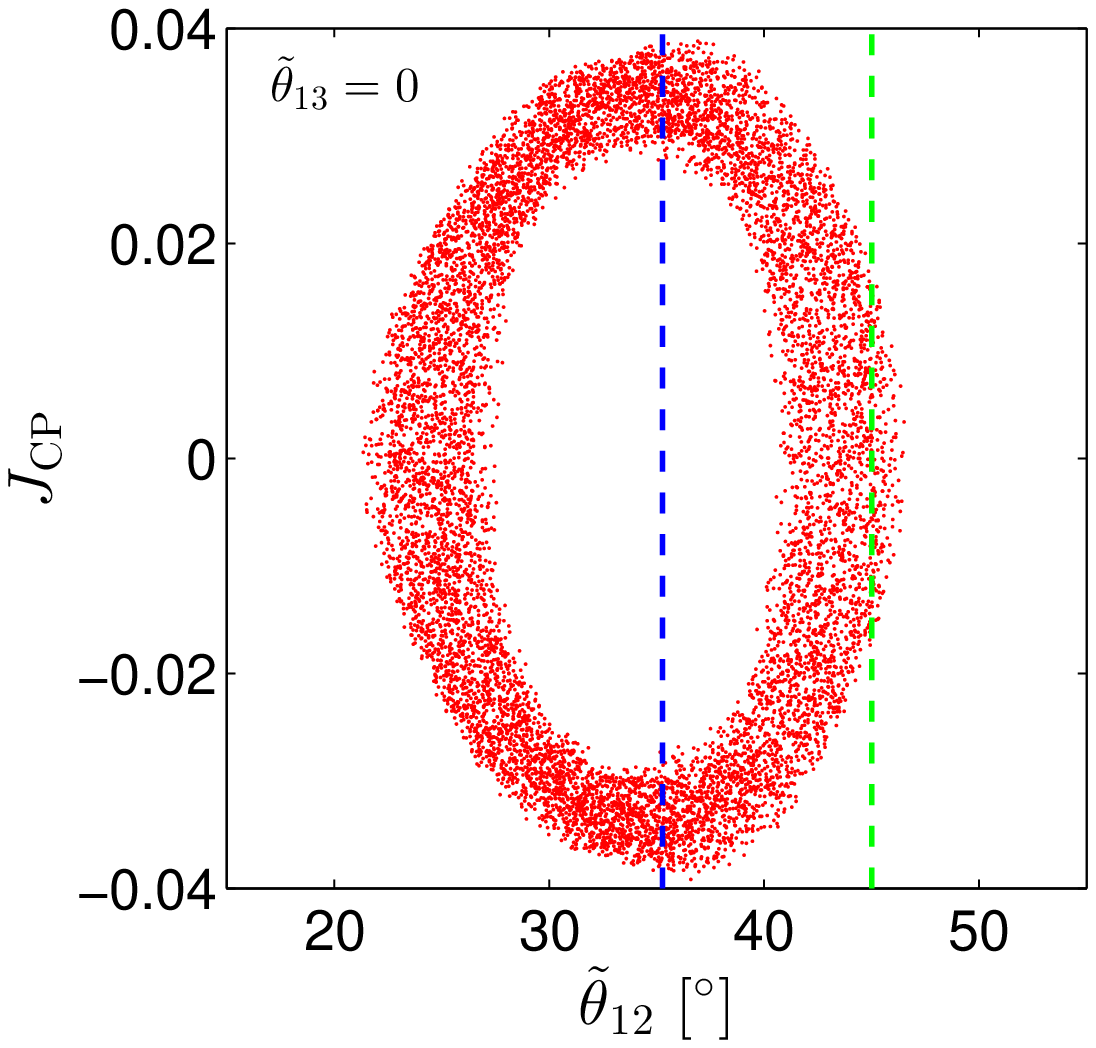} \\
\includegraphics[width=0.4\textwidth]{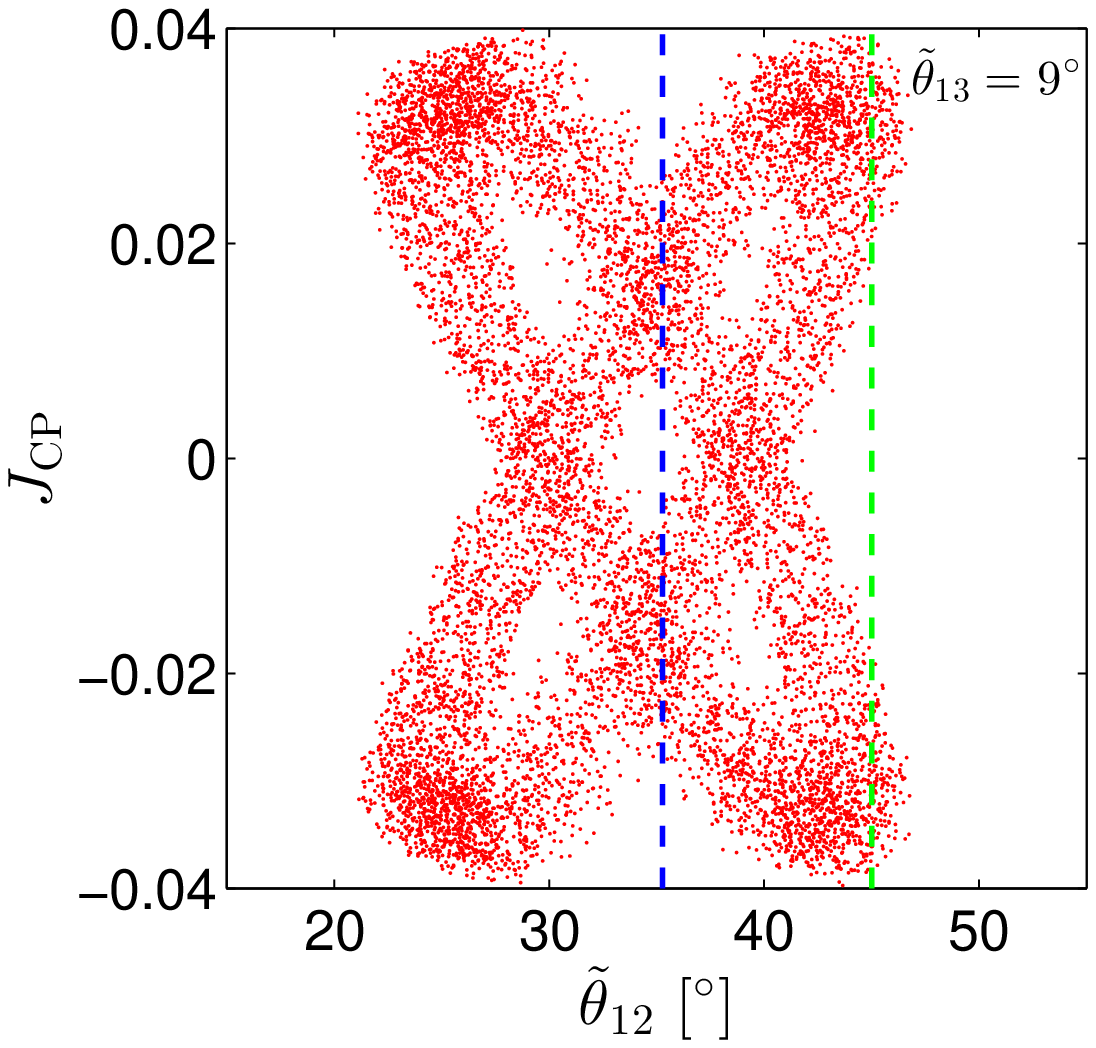}
\includegraphics[width=0.4\textwidth]{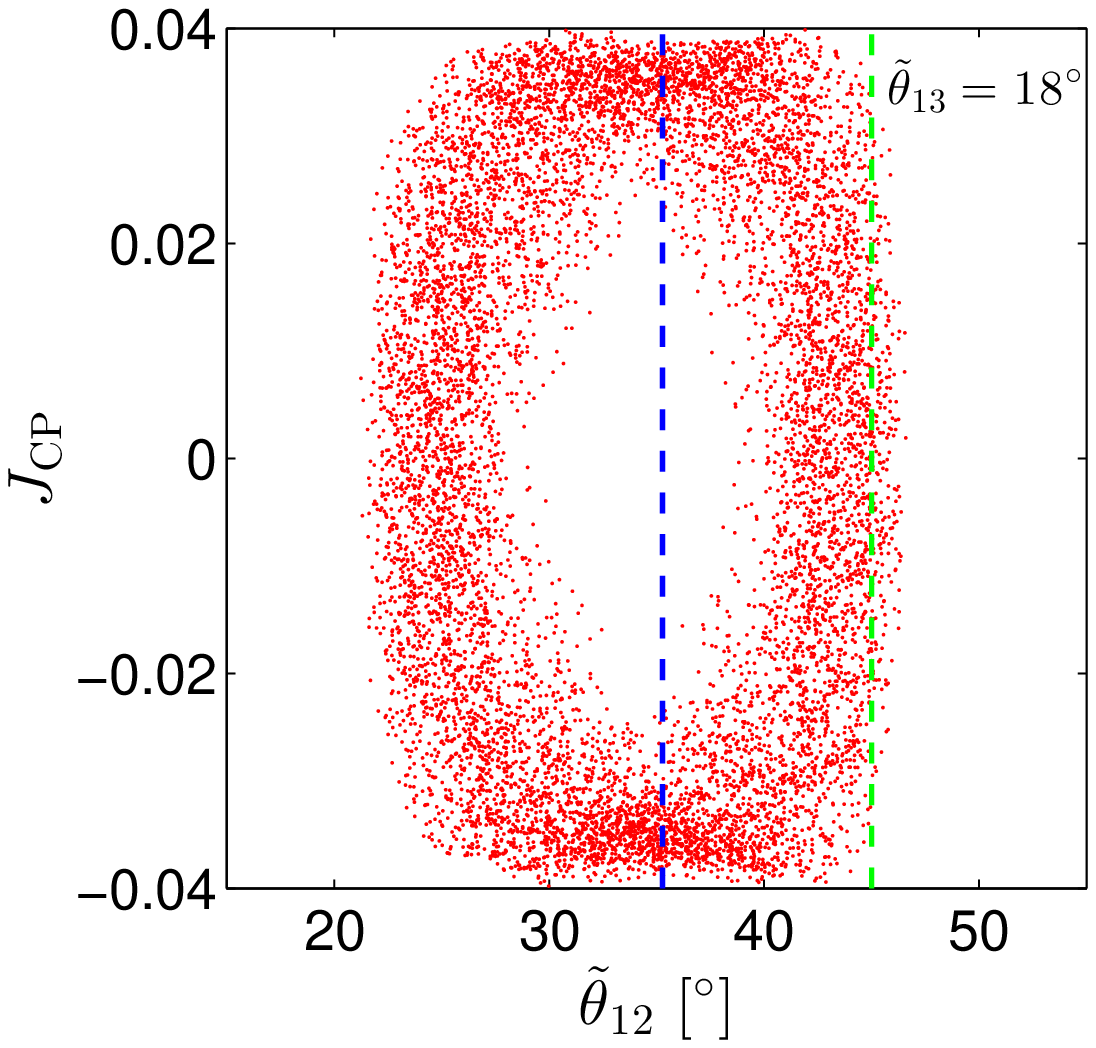}
\caption{\label{fig:JcpT12}Scatter plots for the parameter range of $J_{\rm CP}$
and $\tilde \theta_{12}$ at 3$\sigma$. Here we marginalize all the model parameters for the
upper left plot, and fix $\tilde\theta_{13}=0$, $\tilde\theta_{13}=9^\circ$ and
$\tilde\theta_{13}=18^\circ$ in the other plots, respectively. The blue and green dashed lines
correspond to $\sin^2\tilde\theta_{12}=1/3$ and $\sin^2\tilde\theta_{12}=1/2$.}
\end{figure}

Since the choice of $\tilde \theta_{12}$ can be sensitive to the CP phases, we further
 illustrate in Fig.~\ref{fig:JcpT12} the 3$\sigma$ ranges of the Jarlskog invariant with respect to
$\tilde \theta_{12}$. As one can read from the plot, $J_{\rm CP}$ is not sensitive to $\tilde
\theta_{12}$ in the most general case. However, once $\tilde \theta_{13}$ is fixed, a
connection between $J_{\rm CP}$ and $\tilde \theta_{12}$ can be expected. As we have mentioned in
Sec.~\ref{sec:IIA}, in the case of vanishing $\tilde\theta_{13}$, maximal CP violation ($J_{\rm
CP} \simeq \pm 0.04$) is achieved for $\tilde s^2_{12} \simeq 1/3$ since $\delta$ is
close to $\pi/2$ (or $3\pi/2$). In contrast, $\tilde s^2_{12} \simeq 1/2$ leads to a suppressed
$J_{\rm CP}$ as can be seen from the upper right plot. For the case of $\tilde \theta_{13} \simeq
\theta_{13} \simeq 9^\circ$, our analytical results given in Eq.~\eqref{eq:27} appear as reasonably
good approximations. For instance, the tri-bimaximal value $\tilde s^2_{12} \simeq 1/3$ suggests
$|\sin\delta| \sim 1/2$, corresponding to $J_{\rm CP} \sim J^{\rm max}_{\rm CP}/2$, which is
reflected in the lower left plot. Furthermore, $\tilde s^2_{12} \simeq 1/2$ results in
$|\sin\delta| \sim 0.87$, indicating nearly maximal CP violation. As mentioned above,
the situation for the large $\tilde\theta_{13}$ case is
similar to the $\tilde \theta_{13}$ case, as shown in the discussion
after Eq.~\eqref{eq:25}.

\subsection{Lepton Mixing Parameters}

\begin{figure}[h]
\includegraphics[width=0.4\textwidth]{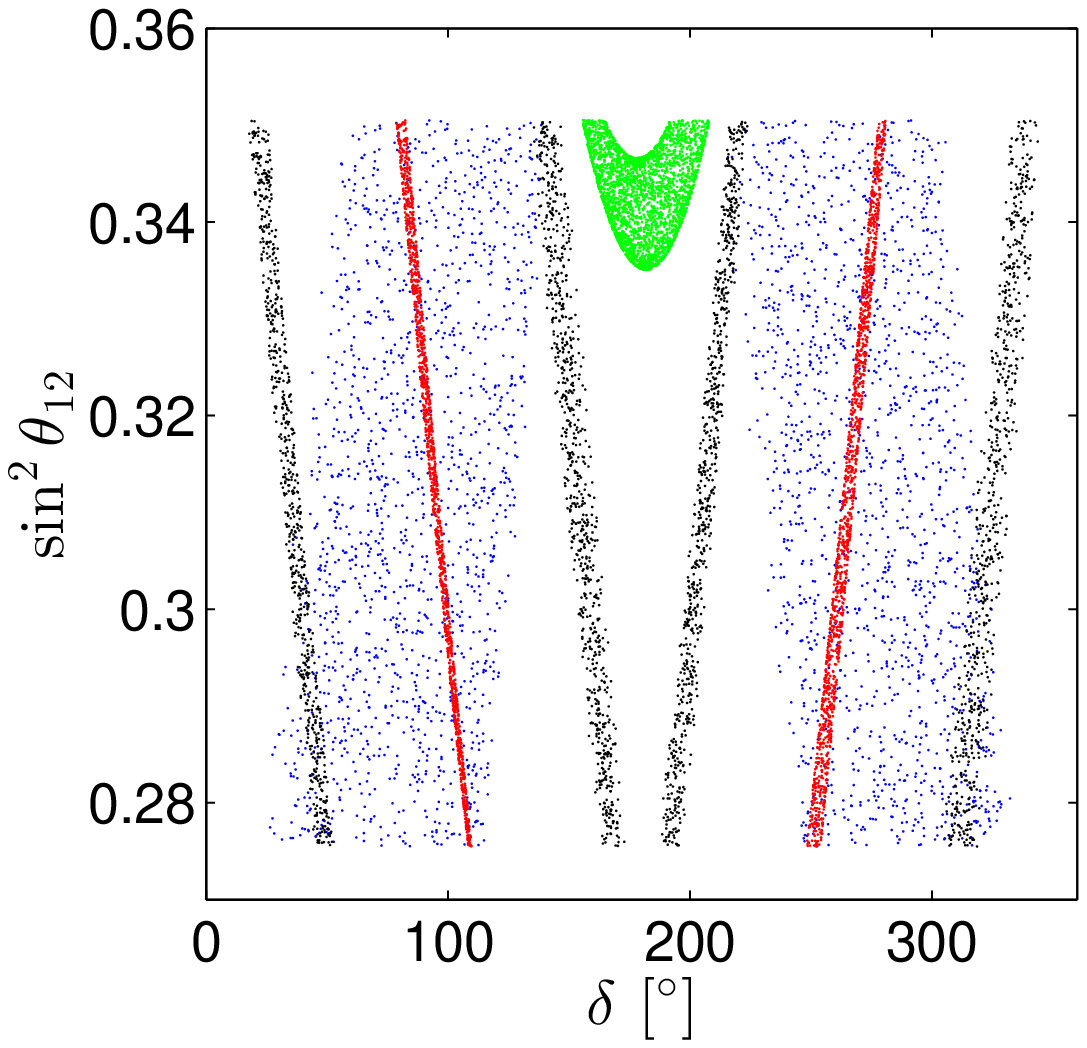}
\includegraphics[width=0.4\textwidth]{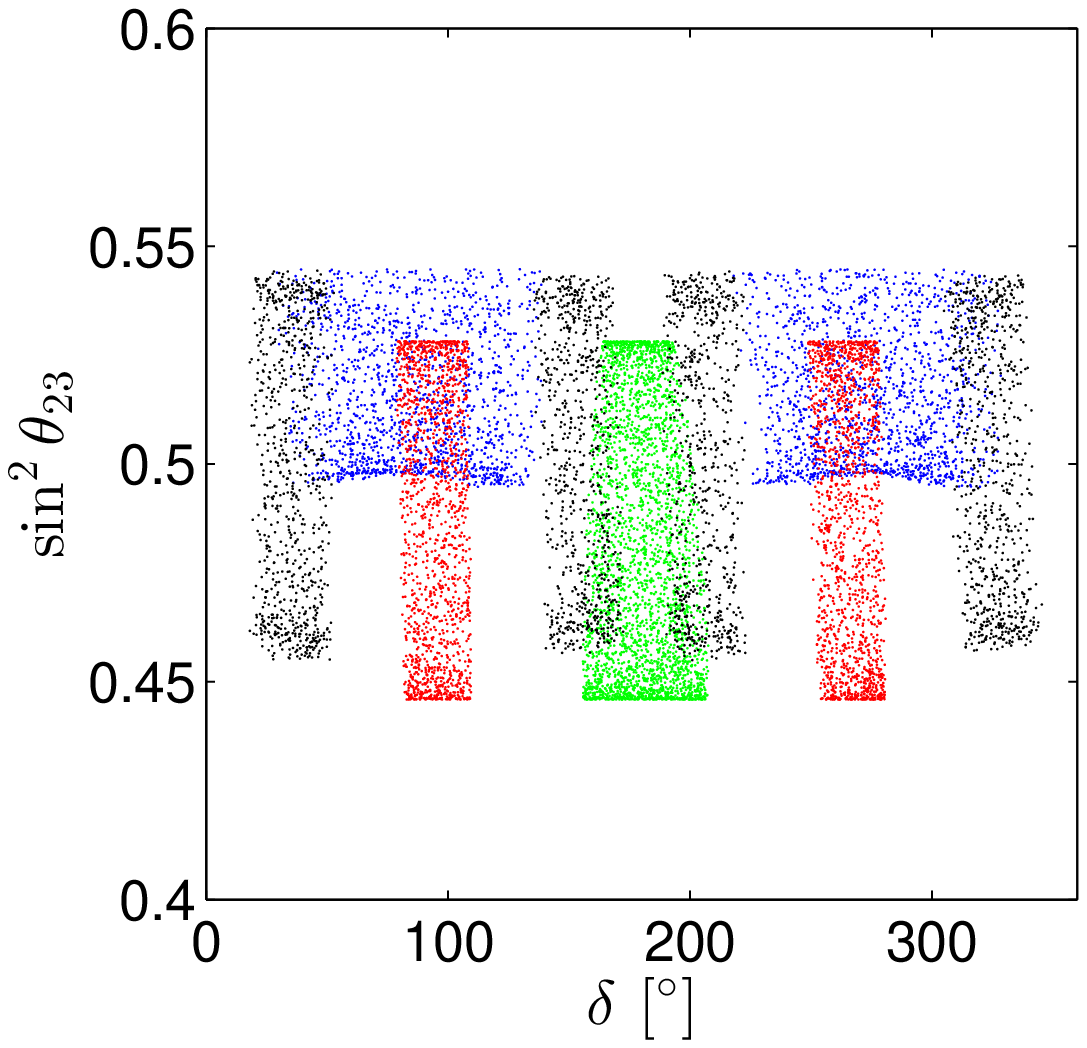}\vspace{-0.0cm} \\
\includegraphics[width=0.4\textwidth]{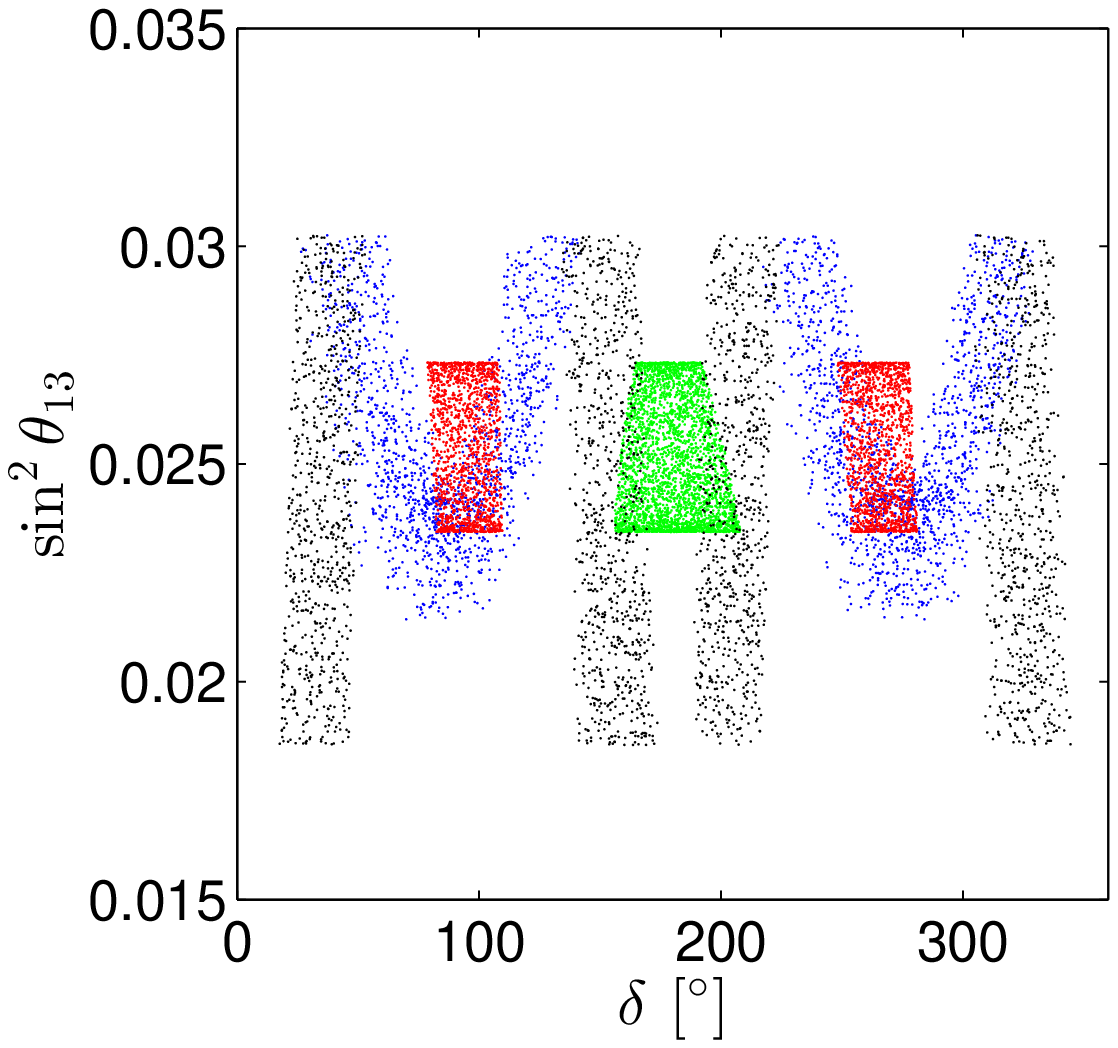}\vspace{-0.0cm}
\includegraphics[width=0.4\textwidth]{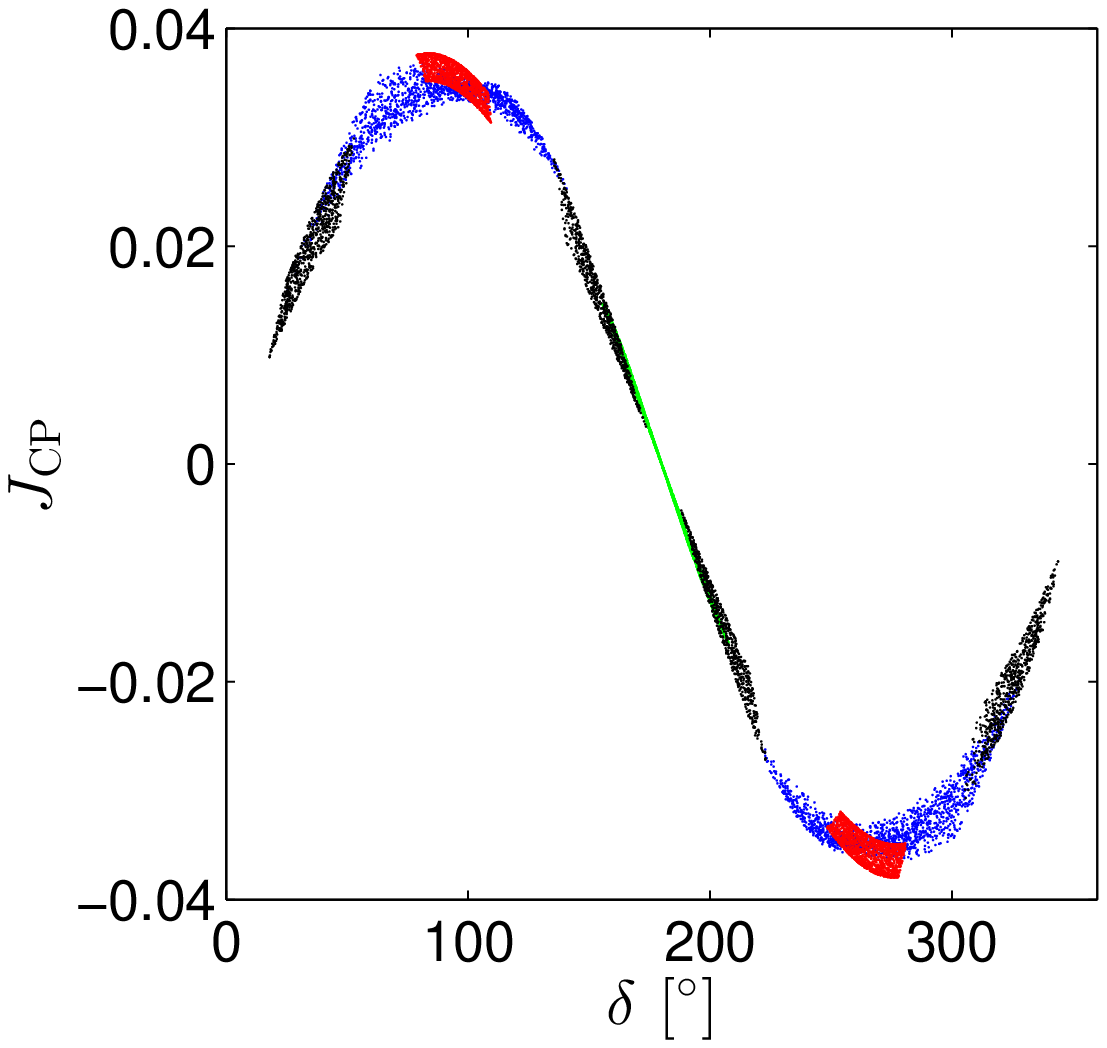}\vspace{-0.0cm} \\
\includegraphics[width=0.4\textwidth]{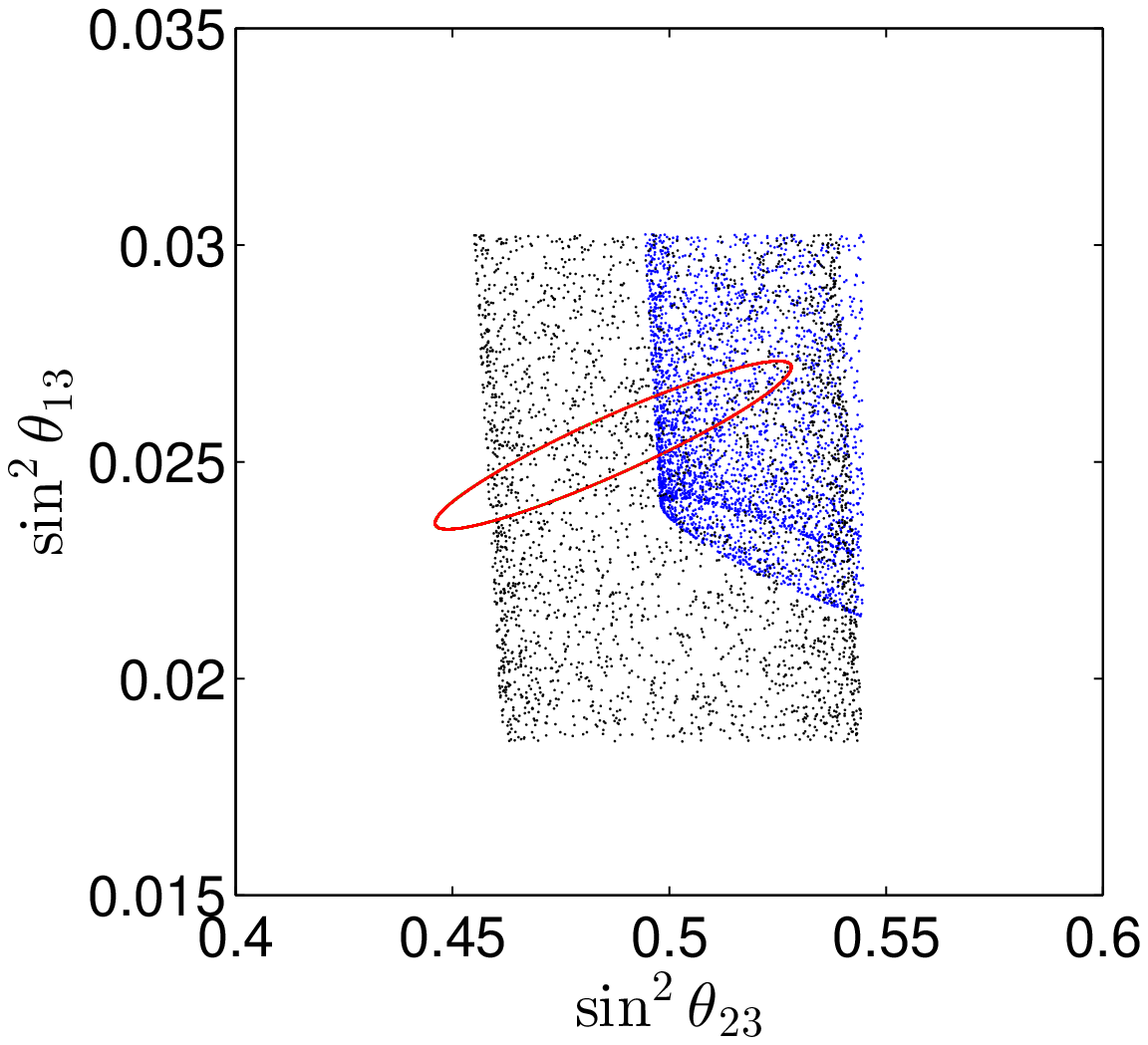}\vspace{-0.0cm}
\includegraphics[width=0.4\textwidth]{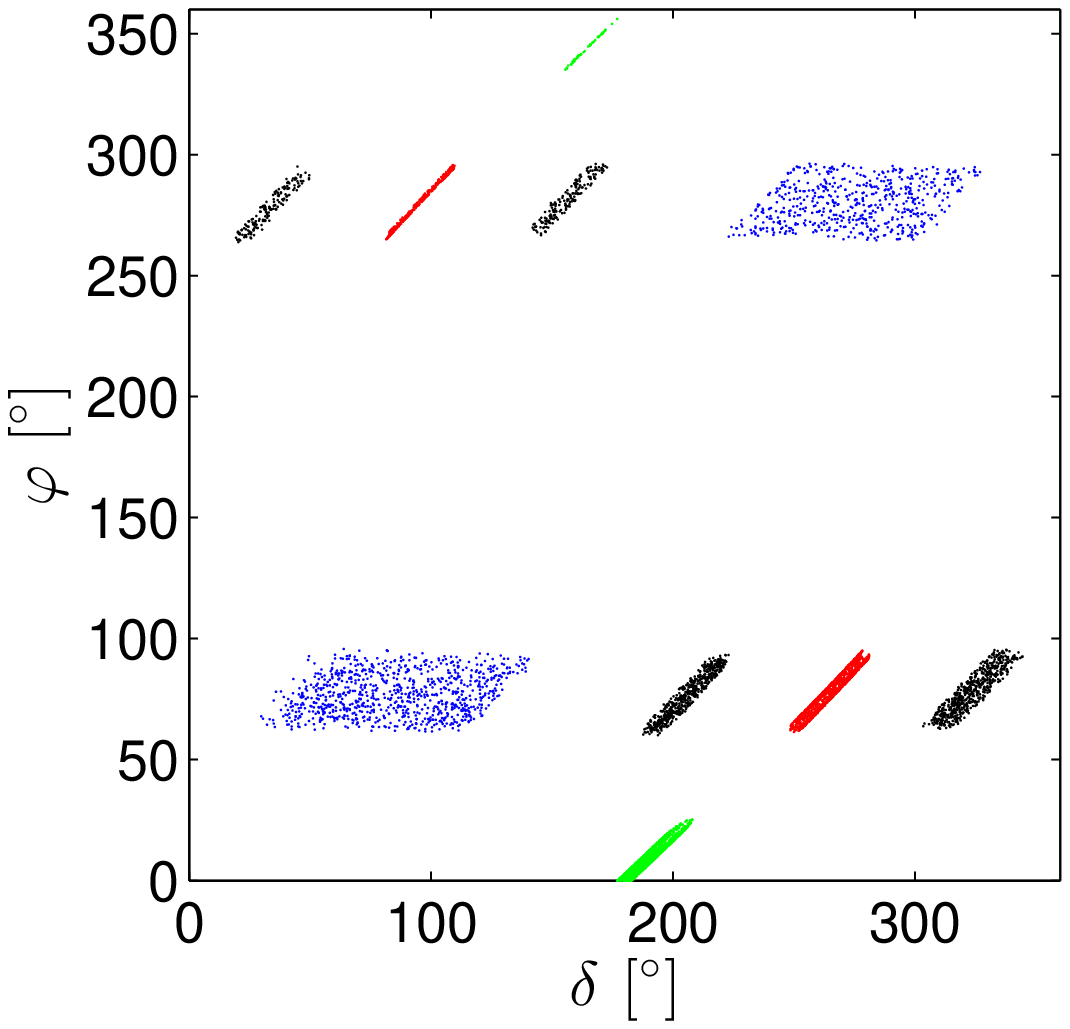}\vspace{-0.0cm}
\caption{\label{fig:5}The allowed 3$\sigma$ range of the lepton mixing parameters and the Jarlskog
invariant. Green (red) points are for bimaximal (tri-bimaximal) mixing in $\tilde U_\nu$. The other
cases are for $\sin^2 \tilde \theta_{12} = 1/3$, $\sin^2 \tilde \theta_{23}=1/2$ and $\tilde
\theta_{13} = \pi/10$ (blue) or $\tilde \theta_{13} = \pi/20$ (black). Since $\theta_{13}$ and $\theta_{23}$ are related in the same way for
cases a) and b), the red and green points are overlapping in the left bottom plot.}
\end{figure}

Finally, the correlations among the leptonic mixing parameters are shown in Fig.~\ref{fig:5}.
We choose four benchmark neutrino mixing matrices $U_{\nu}$: \\
 a) tri-bimaximal pattern with $\tilde \theta_{13}=0$, $\sin^2\tilde \theta_{12}=1/3$ and $\sin^2\tilde \theta_{23}=1/2$ (red points); \\
 b) bimaximal pattern with $\tilde \theta_{13}=0$, $\sin^2\tilde \theta_{12}=1/2$ and $\sin^2\tilde \theta_{23}=1/2$ (green points); \\
 c) large $\tilde \theta_{13}$ case with $\tilde \theta_{13}=\pi/10$, $\sin^2\tilde \theta_{12}=1/3$ and $\sin^2\tilde \theta_{23}=1/2$ (blue points); \\
 d) medium $\tilde \theta_{13}$ case with $\tilde \theta_{13}=\pi/20$, $\sin^2\tilde \theta_{12}=1/3$ and $\sin^2\tilde \theta_{23}=1/2$ (black points); \\
Our analytical results from the previous Sections are confirmed, e.g.,
the tri-bimaximal (bimaximal) pattern leads to $\delta \simeq \pi/2$ ($\delta \simeq \pi$).
When $\tilde \theta_{13}$ is sizable, the Dirac CP
phase depends on $\phi$ and $\varphi$, and therefore is not fixed. However, the choice of
$\varphi$ is restricted from $\theta_{12}$, which in turn sets
constraints on $\delta$.

\newpage
\section{Conclusions}
\label{sec:summary}
Since for a long time only an upper limit on $\theta_{13}$ existed,
most neutrino models were constructed to generate zero $\theta_{13}$. The recent finding of a
sizable value, $\theta_{13} = 9^\circ$, have led to many studies on
generating that value from an initially
zero value. We have noted here that this approach may be misleading, and that in fact $\theta_{13}$ could
have initially been larger. The routinely applied corrections in models will then {\it reduce} $\theta_{13}$
to the observed value, a possibility usually not taken into account. We illustrated the
consequences of this approach in an explicit example based on charged lepton
corrections\footnote{Another approach could be to study radiative corrections to
reduce the value of $\theta_{13}$, or corrections from vacuum misalignment in flavor symmetry models.}.

An extreme case is that initially $\theta_{13}$ corresponds to $18^\circ$, or $\pi/10$. It is
then corrected by $\sin \theta_{\rm C}/\sqrt{2}$ to the observed value of $9^\circ$. Hence, here we
do not have $0 + 9 = 9$, but rather of $18 - 9 = 9$. An analytical and numerical study of the
general case was performed, revealing new correlations and sum rules, different from the usually
considered charged lepton corrections, that are based on initially vanishing $\theta_{13}$.
We find that the correlation of maximal CP violation ($\delta=\pi/2$) for initial tri-bimaximal
mixing and CP conservation ($\delta=\pi$) for initial bimaximal mixing is present for
both extreme cases, initial $\theta_{13} = 18^\circ$ and $\theta_{13} = 0$.

We conclude that the possibility of a more complex mixing pattern than usually considered
should not be ignored. The simple framework studied here is one example where a departure from the
usual approaches results in interesting and novel phenomenology.

\begin{acknowledgments}
This work is supported by the Max Planck Society in the
project MANITOP.
\end{acknowledgments}

\bibliography{bib}

\end{document}